\documentclass[%
superscriptaddress,
 amsmath,amssymb,
 aps, pra,
twocolumn
]{revtex4-2}

\usepackage{graphicx}% Include figure files
\usepackage{dcolumn}% Align table columns on decimal point
\usepackage{bm}% bold math
\usepackage{xcolor}
\usepackage{hyperref}% add hypertext capabilities
\usepackage{booktabs}
\usepackage{array}
\usepackage{longtable}

\def\bra#1{\langle{#1}|}
\def\ket#1{|{#1}\rangle}

\def\ee{{\mathrm{e}}}
\def\ii{{\mathrm{i}}}

\def\dd{{\mathrm{d}}}

\def\bra#1{\langle #1 |}
\def\ket#1{| #1 \rangle}
\def\braket#1#2{\langle #1 | #2 \rangle}

\newcommand{\beq}{\begin{equation}}
\newcommand{\eeq}{\end{equation}}
\newcommand{\barr}{\begin{eqnarray}}
\newcommand{\earr}{\end{eqnarray}}

\def\ii{\mathrm{i}}
\def\dd{\mathrm{d}}

\begin{document}

\title{Spontaneous emission from driven polar quantum systems}

\author{Piotr G\l{}adysz}\email{glad@umk.pl}
 \affiliation{Institute of Physics, Faculty of Physics, Astronomy and Informatics, Nicolaus Copernicus University in Toru\'{n}, Grudziadzka 5/7, 87-100 Toru\'{n}, Poland}
 \affiliation{Dipartimento di Fisica, Universit\`{a} di Bari, I-70126 Bari, Italy}
 
\author{Karolina S\l{}owik}%
 \affiliation{Institute of Physics, Faculty of Physics, Astronomy and Informatics, Nicolaus Copernicus University in Toru\'{n}, Grudziadzka 5/7, 87-100 Toru\'{n}, Poland}
 \affiliation{Institute of Advanced Studies, Nicolaus Copernicus University in Toru\'{n}, ul. Wile\'{n}ska 4, 87-100 Toru\'{n}, Poland}
 
\author{Francesco V. Pepe}
 \affiliation{Dipartimento di Fisica, Universit\`{a} di Bari, I-70126 Bari, Italy}
 \affiliation{INFN, Sezione di Bari, I-70125 Bari, Italy}


\begin{abstract}
We investigate spontaneous radiative processes in a driven polar two-level system whose interaction with the laser field is dominated by broken inversion symmetry rather than by the usual transition dipole coupling. Using a polaron transformation, we derive the dressed eigenstates of the atom–laser system and show that their longitudinal coupling reshapes the spectrum into two displaced harmonic ladders. We then analyze spontaneous transitions induced by a bosonic reservoir, and obtain transition rates that depend on both the laser parameters and the overlap between displaced field states. In the few-photon regime, we identify conditions under which spontaneous emission from the excited state can be strongly suppressed, thereby extending its lifetime, as well as regimes where the ladder structure enables spontaneous absorption from the ground state. In the semiclassical limit of a strong coherent drive, we derive compact analytical expressions for the total transition rates and show that they are governed by Bessel-function weights associated with multiphoton channels. Our results show how broken inversion symmetry qualitatively modifies decay dynamics and radiative cascades, and they establish driven polar quantum systems as a platform for tuning spontaneous light emission beyond the standard inversion-symmetric setting.
\end{abstract}

\maketitle

%\tableofcontents
\section{Introduction}

The dynamic models that describe the interaction between atoms and laser fields constitute a cornerstone of quantum optics, providing information on resonant scattering, resonance fluorescence, and photon antibunching \cite{cohentannoudjiAPI,mollow1969power}. The Rabi model and its rotating-wave counterpart, the Jaynes-Cummings model \cite{shore1993}, provide the basic framework for describing a two-level system (TLS) driven by a single laser mode. For practical purposes, in view of its macroscopic occupation, the laser can be treated as a classical field or, in the dressed-atom approach, as a quantized bosonic mode \cite{cohentannoudji1977dressed, pepe2024dressed}. However, the Rabi model fails to describe polar systems in which the atomic levels involved in the transition are characterized by an intrinsic dipole moment. In addition to their physical relevance, these cases are interesting from the point of view of applications, especially related to the possibility of realizing optically tunable low-frequency radiation sources based on resonantly driven systems \cite{kibis2009,savenko2012,chestnov2017,koppenhofer2016,scala2020}. Broken inversion symmetry of the atomic states, leading to non-vanishing intrinsic dipole moments, has been studied in the context of coherent driving \cite{Avetissian2013,paspalakis2013} and related experiments, involving quantum dots~\cite{Bimberg2009}, dye molecules~\cite{Brode1941}, spin-echo~\cite{Stables1998}, Ramsey interferometry~\cite{MartinezLinares2003}, crystal centers~\cite{Doherty2013,Zhang2018}, and graphene~\cite{Ilani2004,Degen2017}. 
A recent work \cite{Gladysz2025} characterized the impact of intrinsic dipole moments of atomic systems on their coherent interactions with light. 
Ref.~\cite{scala2021beyond} focused on the effects of simultaneous field coupling with both the transition and intrinsic dipole matrix elements on spontaneous emission from an imperfect cavity in the perturbative regime.
In this article, we provide a full quantitative characterization of spontaneous emission dynamics in the specific atom-laser setting, where the only relevant contribution to the atom-laser Hamiltonian is determined by broken inversion symmetry. Though this condition represents an inversion in the usual hierarchy of couplings, where the contribution of intrinsic dipole moments is marginal, it is still physically plausible when the Rabi coupling with the transition matrix element is suppressed by a small projection of the latter on the laser mode polarization. 

\section{\label{subsec:model}Model}
The total Hamiltonian is given as
\begin{equation}
    H = H_0 + H_R + H_{IR},
\label{eq:H_full}
\end{equation}
where $H_0$ describes both TLS, laser, and TLS--laser interaction, $H_R$ 
stands for the bosonic reservoir, and $H_{IR}$ provides TLS--reservoir interaction.

The TLS with its excited $\ket{e}$ and ground $\ket{g}$ states is coupled to a single-mode monochromatic field, which will be referred to as the \textit{laser mode}, as described by the Hamiltonian 
\begin{equation}
    H_0
    = 
    \tfrac{1}{2}\hbar\omega_0 \sigma_z + \hbar\omega_L b^{\dagger} b + \sigma_z \otimes \tfrac{\hbar}{4} ( \Omega_a b^{\dagger} + \Omega^\star_a b).
\label{eq:H_0}
\end{equation}
with $b$ and $b^{\dagger}$ the laser mode annihilation and creation operators, satisfying the canonical commutation relation
\begin{equation}
    [b,b^{\dagger}] = 1 ,
\end{equation}
and
\begin{equation}
    \sigma_z = \ket{e}\bra{e} - \ket{g}\bra{g} .
\end{equation}
The quantities $\omega_0$ and $\omega_L$ represent the TLS transition frequency and the laser frequency, respectively. The coupling between TLS and field emerges from an interaction Hamiltonian in the form $-\vec{d}\cdot\vec{E}(0)$ in the dipole representation \cite{cohen2_1998}, where $\vec{d}$ is the TLS dipole moment and $\vec{E}(0)$ is the field operator evaluated in the TLS position. It is worth noticing that, if the field propagates in a dispersive medium, it generally involves both transverse and longitudinal components \cite{scala2020}. As shown in Appendix~\ref{app:model}, the coupling strength can be expressed in terms of microscopic quantities as
\begin{equation}\label{eq:Omegaa}
    \Omega_a = -2 (\vec{d}_{e} - \vec{d}_g)\cdot\vec{\epsilon}(0) \sqrt{\frac{\omega_L}{2\hbar\varepsilon_0}} \varphi(0)^* ,
\end{equation}
where $\varphi(0)$ and $\vec{\epsilon}(0)$ are the mode amplitude and polarization, evaluated at the TLS position, while $\vec{d}_{e} = \bra{e} \vec{d} \ket{e}$ and $\vec{d}_{g} = \bra{g} \vec{d} \ket{g}$ are the intrinsic dipole moments of the two levels. The intrinsic dipoles and the transition dipole moment must be regarded as generally independent quantities. In particular, $\vec{d}_{eg} = \bra{e} \vec{d} \ket{g}$ crucially depends on the spatial overlap of the two wavefunction, whereas $\vec{d}_e$ and $\vec{d}_g$ obviously do not.

The TLS is further submerged in a bosonic reservoir with free Hamiltonian 
\begin{equation}
    H_R
    = 
    \int \dd \vec{k} \sum_\epsilon \hbar\omega(\vec{k}) a^\dagger_\epsilon(\vec{k}) a_\epsilon(\vec{k}),
\label{eq:H_R}
\end{equation}
where $a_\epsilon(\vec{k})$ describes reservoir modes of polarization $\epsilon$, wave vector $\vec{k}$ and frequency $\omega(\vec{k})$. The interaction of the TLS with the reservoir is dominated by the transverse coupling
\begin{equation}
    H_{IR} 
    = 
    \hbar \int \dd \vec{k} \sum_\epsilon \sigma_x \otimes\left( g_\epsilon(\vec{k}) \, a^\dagger_\epsilon(\vec{k}) + g^\star_\epsilon(\vec{k}) \, a_\epsilon(\vec{k}) \right) 
\label{eq:H_IR}
\end{equation}
of strength $g_\epsilon(\vec{k})$. Here, $\sigma_x = \ket{e}\bra{g}+\ket{g}\bra{e}$ describes the TLS transitions. We neglect longitudinal system-reservoir interaction but include the full form of the transverse coupling without the typically used rotating-wave approximation. Such an approximation is legitimate as long as the analysis is focused on perturbative decay dynamics towards free-space modes. Actually, terms proportional to $\sigma_z\otimes a_\epsilon(\vec{k})$ and $\sigma_z\otimes a^{\dagger}_\epsilon(\vec{k})$ cannot produce on-shell final states, and thus do not contribute to the decay rate at the lowest order in the coupling constant.

\subsection{\label{subsec:polaron}Polaron transformation}
We adopt the unitary polaron transformation \cite{wilson2002}
\begin{equation}
    U 
    = 
    \exp\left(\sigma_z \otimes \left( \tfrac{\Omega_a}{4\omega_L} b^\dagger - \tfrac{\Omega_a^\star}{4\omega_L} b \right)\right)
\label{eq:polaron}
\end{equation}
to remove the explicit longitudinal coupling term from Eq.~\eqref{eq:H_0} and find an effective form of the interaction in a dressed-states representation. 
The transformation operator in Eq.~\eqref{eq:polaron} can be rewritten in the form of field displacement $D (\alpha)=\exp\bigl(\alpha b^\dagger-\alpha^\star b\bigr)$ conditional on the TLS state
\begin{equation}
    U 
    = 
    \ket{e}\bra{e} \otimes D \left(\tfrac{\Omega_a}{4\omega_L} \right)
    + \ket{g}\bra{g} \otimes D \left(-\tfrac{\Omega_a}{4\omega_L} \right),
\label{eq:splitPolaron}
\end{equation}
The transformed Hamiltonian is then given by $\tilde{H} = UHU^\dagger$. In the transformed frame, the terms given by Eqs.~\eqref{eq:H_0} --~\eqref{eq:H_IR} read respectively

\begin{subequations}
    \label{eq:tildeH}
    \begin{align}
    &\tilde{H}_0 / \hbar
    = 
    \tfrac{\omega_0}{2}\sigma_z + \omega_L b^\dagger b - \tfrac{|\Omega_a|^2}{16 \omega_L}, 
    \label{eq:tildeH_0}
    \\
    &\tilde{H}_R / \hbar
    = 
    H_R/\hbar, 
    \label{eq:tildeH_R}
    \\
    \begin{split}
        &\tilde{H}_{IR} / \hbar
        = 
        \left( D\left( \tfrac{\Omega_a}{2\omega_L} \right) \sigma_+ + D\left( -\tfrac{\Omega_a}{2\omega_L} \right) \sigma_- \right) 
        \\
        & \qquad \qquad \otimes \int \dd \vec{k} \sum_\epsilon \left( g_\epsilon(\vec{k}) \, a^\dagger_\epsilon(\vec{k}) + g^\star_\epsilon(\vec{k}) \, a_\epsilon(\vec{k}) \right).
    \end{split}
    \label{eq:tildeH_IR}
    \end{align}
\end{subequations}
In Eq.~\eqref{eq:tildeH_0}, the constant term trivially shifts the total energy and can be dropped. The free reservoir term Eq.~\eqref{eq:tildeH_R} is unchanged as no operators act on the reservoir's Hilbert space in the transformation. The system-reservoir interaction Eq.~\eqref{eq:tildeH_IR} is modified in the presence of the laser field; $\sigma_+=\sigma_-^\dagger = \ket{e}\bra{g}$ are atomic ladder operators.
Appendix \ref{apx:polaron} provides detailed derivations of Eqs.~\eqref{eq:splitPolaron}, and~\eqref{eq:tildeH}.

\subsection{\label{subsec:eigen}Eigenstates}
\begin{figure}[t]
    \centering
    \includegraphics[width=0.99\linewidth]{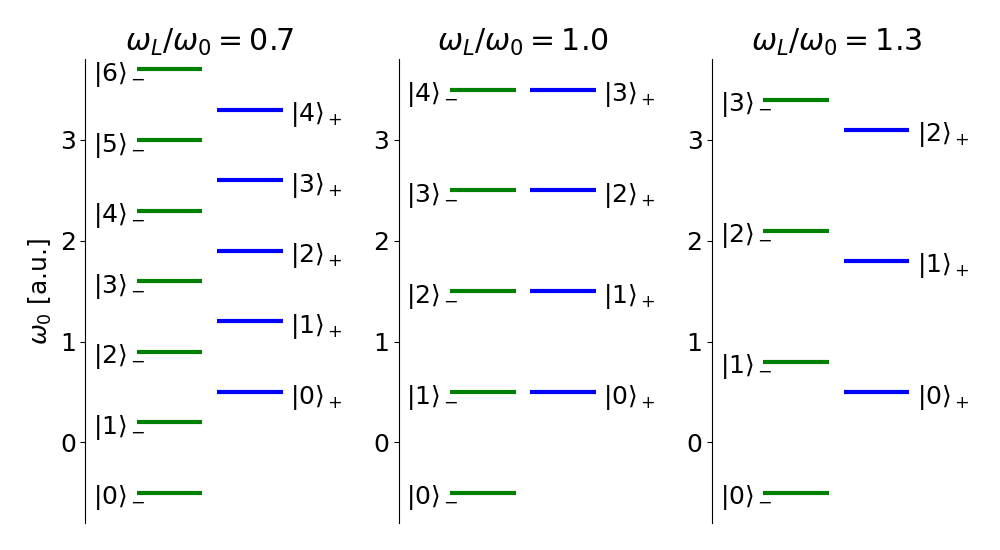}
    \caption{Energy levels of the untransformed system interacting with the laser field via longitudinal coupling. Green lines (left ladder in each plot) correspond to the system's ground state, while blue ones (right ladder in each) - to the system's excited state. Note that increasing $\omega_L$ "stretches" the ladders while the difference between the lowest levels stays the same.}
    \label{fig:energy_levels}
\end{figure}

The transformed free Hamiltonian $\tilde{H}_0$ separates the atomic and field Hilbert spaces, hence the transformed eigenstates can be trivially written as 
$\ket{\tilde{\psi}_{i,n}} \equiv \ket{i}\otimes\ket{n}$,
where 
$i \in \{e,g\}$,
and 
$\ket{n}$
is the Fock state such that
$b^\dagger b \ket{n} 
=
n \ket{n}$. 
Thus, we have 
$\tilde{H}_0/ \hbar \ket{\tilde{\psi}_{i,n}} 
=
\big(s_i \tfrac{\omega_0}{2} + n\omega_L\big) \ket{\tilde{\psi}_{i,n}}$,
where sign 
$s_i 
=
+,\,-$ for $i=e,\,g$,
respectively.
The eigenstates of the original $H_0$ Hamiltonian (in the laboratory frame) are then easily obtainable as 
$\ket{\psi_{i,n}} = U^\dagger \ket{\tilde{\psi}_{i,n}}$.
Due to the form of the unitary transformation \eqref{eq:splitPolaron}, the laser states now depend on the atomic state
\begin{equation}
    \ket{\psi_{i,n}} 
    = 
    \ket{i}\otimes D\left( -s_i \tfrac{\Omega_a}{4\omega_L} \right) \ket{n} \equiv\ket{i} \otimes \ket{n}_{s_i},
\label{eq:psi_in}
\end{equation}
where $\ket{n}_{s_i}$ are displaced Fock states denoted by \mbox{$\ket{-s_i\tfrac{\Omega_a}{4\omega_L},n}$}. The laser ground states are also coherent states since for $n=0$ we have $\ket{0}_\pm = D\big( \mp \tfrac{\Omega_a}{4\omega_L} \big) \ket{0} \equiv \ket{\mp \tfrac{\Omega_a}{4\omega_L}}$ in the original basis. For the same atomic state, the laser states are orthonormal ${}_\pm\braket{\ell}{n}_\pm = \delta_{\ell n}$; otherwise, they overlap according to the expression 
\begin{equation}
    \begin{split}
    {}_\pm\braket{\ell}{n}_\mp 
    =&
    (\pm)^\ell (\mp)^n \sqrt{\ell! \, n!} \; \ee^{-\tfrac{1}{2}\left| \tfrac{\Omega_a}{2\omega_L} \right|^2} \ee^{\ii (\ell - n)\phi } 
    \\
    &\times \sum_{k=0}^{\mathrm{min}(\ell, n)} \tfrac{(-1)^k}{k! (\ell - k)! (n - k)!} \left| \tfrac{\Omega_a}{2\omega_L} \right|^{\ell + n -2k},
    \end{split}
\label{eq:ln_overlap}
\end{equation}
where $\phi = \arg(\Omega_a)$ arises from the complex nature of the coupling and does not affect the overall physics discussed further, as all the observables depend on the absolute value of the states overlap. As a result, the longitudinal coupling to the drive turns the TLS+field energy configuration into two split harmonic-oscillator ladders, one for the ground and one for the excited TLS state, shifted by the energy $\omega_0$ (Fig.~\ref{fig:energy_levels}). States are orthogonal between the ladders since $\langle g|e\rangle=0$, even if the field states overlap, while states within the same ladder do not overlap. 
In the lab frame, the field eigenstates are Fock states displaced according to the strength of the longitudinal coupling $\Omega_a$. This is a particularly important observation because exciting a specific eigenstate in the system requires the driving field to be precisely tuned to the coupling strength, e.g., prepared in the coherent state $\ket{-\tfrac{\Omega_a}{4\omega_L}}\equiv\ket{0}_+$. Appendix \ref{apx:eigenstates} provides detailed description and derivation of the overlapping function \eqref{eq:ln_overlap}.

\section{\label{subsec:spontaneous}Spontaneous emission}
To analyze the influence of the longitudinal coupling onto the system-reservoir interaction we use the resolvent theory \cite{cohen2_1998}. We are interested in the initial atom-laser state given by Eq.~\eqref{eq:psi_in} and the reservoir in the vacuum state $\ket{\text{vac}}$. The full, untransformed initial state for the total Hamiltonian $H$ is then
$\ket{\Psi_{i,n}} = \ket{\psi_{i,n}}\otimes\ket{\text{vac}}$,
while in the rotated frame for $\tilde{H}$ we have decoupled TLS and laser's modes along with untouched reservoir vacuum state
\begin{equation}
    \ket{\tilde{\Psi}_{i,n}} 
    = 
    \ket{\tilde{\psi}_{i,n}} \otimes \ket{\text{vac}} \equiv \ket{i} \otimes \ket{n} \otimes \ket{\text{vac}},
\label{eq:tildePsi_in}
\end{equation}

\begin{figure*}[ht!]
    \centering
    \includegraphics[width=0.91\linewidth]{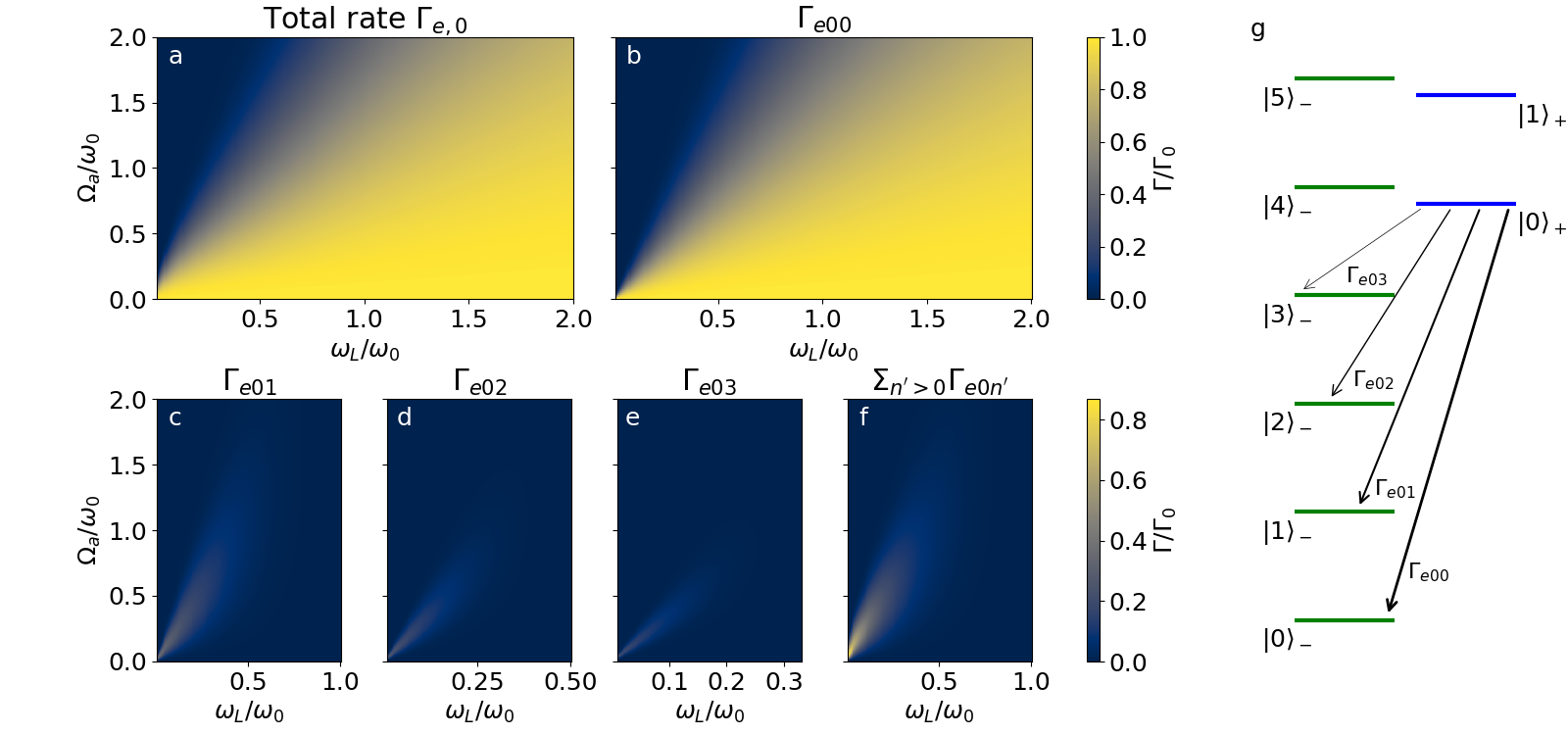}
    \caption{Spontaneous emission rates for the initial state $\ket{\Psi_{e,0}}$ normalized to $\Gamma_0$ as functions of coupling strength $\Omega_a$ and laser frequency $\omega_L$. Colorbar in each row corresponds to the plots in that row.
    (a) Total rate including all the possible transitions to the lower states.
    (b)--(e) Partial rates for transitions to states $\ket{g} \otimes \ket{n^\prime}_-$ for $n^\prime=0, \, 1, \, 2, \,\text{and} \; 3$. 
    (f) Sum of partial rates for $n^\prime>0$ -- multiphoton correction.
    (g) Scheme of the ladder structure for a strongly red-detuned case (see Fig.~\ref{fig:energy_levels} for reference), with arrows marking the states connected by the transition shown in each subplot. Arrow thickness indicates the importance of the process to the overall emission from most important (thick) to least important (thin).}
    \label{fig:Gamma_e0}
\end{figure*}

In the transformed frame the resolvent for such initial state Eq.~\eqref{eq:tildePsi_in} reads
\begin{equation}
    \begin{split}
    G_{i,n}(z) 
    &= 
    \bra{\tilde{\Psi}_{i,n}} \tfrac{1}{z - \tilde{H}} \ket{\tilde{\Psi}_{i,n}}
    \\
    &= 
    \left(z - \bra{\tilde{\Psi}_{i,n}} \left( \tilde{H}_0 + \tilde{H}_R \right) \ket{\tilde{\Psi}_{i,n}} - \Sigma_{i, n}(z)\right)^{-1},
    \end{split}
\label{eq:resolvent}
\end{equation}
where $\Sigma_{i,n}(z)$ is the self energy term. In the weak system-reservoir interaction regime it can be written as
\begin{equation}
    \Sigma_{i, n}(z)
    \approx 
    \bra{\tilde{\Psi}_{i,n}} \tilde{H}_{IR} \tfrac{1}{z - \tilde{H}_0 - \tilde{H}_R} \tilde{H}_{IR} \ket{\tilde{\Psi}_{i,n}},
\label{eq:generalSelfEnergy}
\end{equation}
with $z$ being the complex variable corresponding to the quasi-energy at which the self energy is evaluated. Evaluating it at 
$z_{i,n} = s_i\tfrac{\omega_0}{2} + n\omega_L + \ii0$,
corresponding to the energy of the initial state $\ket{\tilde{\Psi}_{i,n}}$, yields the spectral shift $\Delta_{i,n}$ and the broadening $\Gamma_{i,n}$ of this energy level. In the case of the longitudinally driven TLS, one finds
\begin{equation}
    \Sigma_{i,n}\left( z_{i,n} \right) 
    = \Delta_{i,n} + \tfrac{\ii}{2} \Gamma_{i,n} = 
    \sum_{n^\prime} \left( \Delta_{inn^\prime} - \tfrac{\ii}{2} \Gamma_{inn^\prime} \right).
\label{eq:selfEnergy}
\end{equation}
The terms $\Delta_{inn^\prime}$ and $\Gamma_{inn^\prime}$ have an intuitive interpretation as they describe partial corrections arising from the allowed transitions 
$\ket{\tilde{\Psi}_{i,n}} \rightarrow \ket{j} \otimes \ket{n^\prime} \otimes \ket{1_{\epsilon(\vec{k})}}$. 
The form of TLS-reservoir interaction Hamiltonian Eq.~\eqref{eq:tildeH_IR} prevents transitions within the same energy ladder, hence $i\neq j$. Moreover, only transitions with emission of a single photon $\ket{1_{\epsilon(\vec{k})}}$ to the reservoir are included due to the assumption of weak system-reservoir interaction. As the frequency of the emitted photon has to be positive, the following restriction holds: 
\begin{equation}
    n^\prime \leq n + s_i \tfrac{\omega_0}{\omega_L},
\label{eq:nRestriction}
\end{equation}
meaning that spontaneous transitions only occur to states energetically below the initial one and between the energy ladders (see Appendix~\ref{apx:resolvent} for details). In consequence, the summation over $n^\prime$ in Eq.~\eqref{eq:selfEnergy} is determined individually for the initial state and the shape of the energy ladders (defined by $\omega_0$ and $\omega_L$ frequencies; see Fig.~\ref{fig:energy_levels}). Note also that since, in general, $n\neq n^\prime$, these partial terms may describe transitions for which the number of photons in the driving field is modified by more than 1.
In free space, the expression for the $\Gamma_{inn^\prime}$ rates reads
\begin{equation}
    \Gamma_{inn^\prime} /\Gamma_0 
    =
    \left| {}_{s_i}\braket{n}{n^\prime}_{-s_i} \right|^2 \left( s_i1 + (n-n^\prime) \tfrac{\omega_L}{\omega_0} \right)^3,
\label{eq:partial_gamma}
\end{equation}
where $\Gamma_0$ stands for the spontaneous emission rate between the excited and ground states of a TLS in the absence of longitudinal coupling, when the ladder structure collapses back to the standard two energy levels. Appendix~\ref{apx:resolvent} contains details behind the derivation of Eqs.~\eqref{eq:generalSelfEnergy} --~\eqref{eq:partial_gamma} along with the expression of $\Gamma_0$.

\subsection{Single-photon spontaneous-emission suppression}
We now consider the case of a laser drive in the precisely chosen coherent state $\ket{0}_+$.
The only possible transitions are $\ket{e} \otimes \ket{0}_+ \rightarrow \ket{g} \otimes \ket{n^\prime}_-$, with the emission of a single photon to the reservoir. According to Eq.~\eqref{eq:partial_gamma}, the rates for these processes are
\begin{equation}
    \Gamma_{e0n^\prime} / \Gamma_0
    =
    \ee^{-\left| \tfrac{\Omega_a}{2\omega_L} \right|^2} \tfrac{1}{n^\prime!} \left| \tfrac{\Omega_a}{2\omega_L} \right|^{2n^\prime} \left( 1 - n^\prime \tfrac{\omega_L}{\omega_0} \right)^3,
\label{eq:gamma_e0n}
\end{equation}
where Eq.~\eqref{eq:ln_overlap} was used to evaluate the ${}_+\braket{0}{n^\prime}_-$ term. Thus, summing Eq.~\eqref{eq:gamma_e0n} over all allowed $n^\prime$s, the total spontaneous emission rate from the excited state reads
\begin{equation}
    \Gamma_{e,0} / \Gamma_0 
    =
    \ee^{-\left| \tfrac{\Omega_a}{2\omega_L} \right|^2} 
    \sum_{n^\prime=0}^{\lfloor \omega_0/\omega_L \rfloor} \tfrac{1}{n^\prime!} \left| \tfrac{\Omega_a}{2\omega_L} \right|^{2n^\prime} \left( 1 - n^\prime \tfrac{\omega_L}{\omega_0} \right)^3
    \leq 1,
\label{eq:gamma_e0}
\end{equation}
where $\lfloor \omega_0/\omega_L \rfloor$ is a floor function returning the largest integer below the frequency ratio [based on Eq.~\eqref{eq:nRestriction}]. For each term in the above sum, the cubic bracket is non-negative and bounded by 1, hence the sum is upper-bounded by $\exp{| \tfrac{\Omega_a}{2\omega_L} |^2}$, with equality at $\omega_L \rightarrow 0$. Hence, the laser drive suppresses spontaneous emission and enhances the excited-state lifetime. The process is most effective in the red-detuned regime (see Fig.~\ref{fig:Gamma_e0}). The suppression increases with longitudinal coupling $\Omega_a$ [see Fig.~\ref{fig:Gamma_e0}a showing Eq.~\eqref{eq:gamma_e0}]. For the blue-detuned case of $\omega_L > \omega_0$ the floor function returns $\lfloor \omega_0/\omega_L \rfloor = 0$, hence only $n^\prime=0$ is allowed (Fig.~\ref{fig:Gamma_e0}b). Otherwise, additional transitions are unlocked (Fig.~\ref{fig:Gamma_e0}c-f). 
Note that the stepwise structure introduced by floor function is observable via appearance of additional peaks after crossing a threshold. However, the peak smoothly grows with its amplitude scaled with $\omega_L$, due to the cube form factor and so no step is observed in $\Gamma_{e,0}$.

The spontaneous-emission suppression discussed above occurs for the driving field being carefully tuned to the longitudinal coupling strength. Otherwise, the initial state contains other eigenstates of the form $\ket{e} \otimes \ket{n}_+$ which contribute to the overall emission. Note also that switching off the longitudinal coupling $\Omega_a\rightarrow 0$ or the field frequency $\omega_L \rightarrow 0$ reproduces the standard TLS spontaneous emission rate $\Gamma_{e,0}\rightarrow \Gamma_0$.

\subsection{Single-photon spontaneous absorption}

\begin{figure}[t!]
    \centering
    \includegraphics[width=0.99\linewidth]{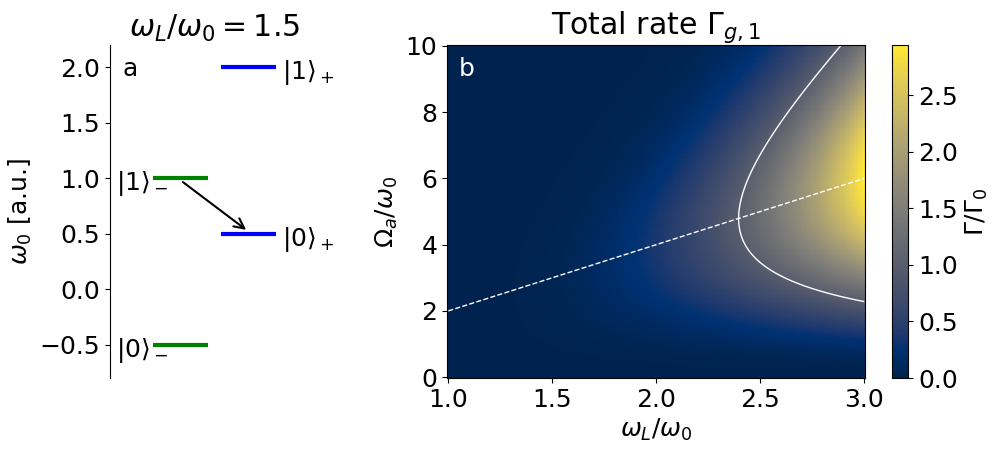}
    \caption{ Spontaneous absorption.
    (a) Energy ladders (green for $\ket{g}$, and blue for $\ket{e}$ TLS states) with possible transition leading to spontaneous absorption marked with arrow. 
    (b) Spontaneous absorption rate for the initial state $\ket{\Psi_{g,1}}$ normalized to $\Gamma_0$ as function of coupling strength $\Omega_a$ and laser frequency $\omega_L$. Dashed line marks maximum values for given frequency, while solid line marks $\Gamma_{g,1}/\Gamma_0=1$ contour.}
    \label{fig:Gamma_g1}
\end{figure}

As spontaneous absorption, we understand the process in which the TLS transfers from its ground to the excited state, while the hybrid system of TLS+field goes down between the energy ladders (see Fig.~\ref{fig:Gamma_g1}a). This class of processes originates from the counter-rotating $\ket{e}\bra{g}\otimes a^\dagger_\epsilon(\vec{k})$ terms in the TLS-reservoir interaction Hamiltonian. The fundamental transition of this kind $\ket{g} \otimes \ket{1}_- \rightarrow \ket{e} \otimes \ket{0}_+$ occurs for the blue-detuned case of  $\omega_L > \omega_0$. Then, based on Eq.~\eqref{eq:partial_gamma}
\begin{equation}
    \Gamma_{g10} / \Gamma_0 
    =
    \ee^{-\left| \tfrac{\Omega_a}{2\omega_L} \right|^2} \left| \tfrac{\Omega_a}{2\omega_L} \right|^2 \left( \tfrac{\omega_L}{\omega_0} - 1 \right)^3.
\label{eq:gamma_g10}
\end{equation}
The result is shown in Fig.~\ref{fig:Gamma_g1}b. For a fixed drive frequency $\omega_L$, the absorption rate is maximized for $\Omega_a = 2\omega_L$ (dashed white line), and grows indefinitely with $\omega_L$. This transition rate can even exceed the spontaneous emission rate $\Gamma_0$ in the absence of a laser mode (solid white line). This would occur due to the involvement of higher-order processes via the counter-rotating part of the system-reservoir interaction. The efficiency of this process grows cubically with the laser detuning from the TLS transition. 

In general, the energy diagram of $H_0$ allows for transitions of the kind
\begin{equation}
    \ket{g} \otimes \ket{n}_- \rightarrow \ket{e} \otimes \ket{n'}_+ ,
\end{equation}
occurring whenever $(n-n')\omega_L > \omega_0$, which are made possible by the fact that, due to the presence of a dominant longitudinal coupling in the TLS-laser Hamiltonian, the overlap of $\ket{n}_-$ and $\ket{n'}_+$ is nonvanishing. These processes have no analogue in the Jaynes-Cummings Hamiltonian, where occurs from eigenstates $\ket{\psi}=\alpha\ket{e}\otimes\ket{n} + \beta\ket{g}\otimes\ket{n+1}$ towards $\ket{\psi'}=\alpha'\ket{e}\otimes\ket{n-1} + \beta'\ket{g}\otimes\ket{n}$, and therefore gets no contribution from the $\ket{e}\bra{g}\otimes a^\dagger_\epsilon(\vec{k})$ terms of the TLS-reservoir interaction.

\subsection{Semiclassical limit of spontaneous emission}
\begin{figure}[t!]
    \centering
    \includegraphics[width=0.9\linewidth]{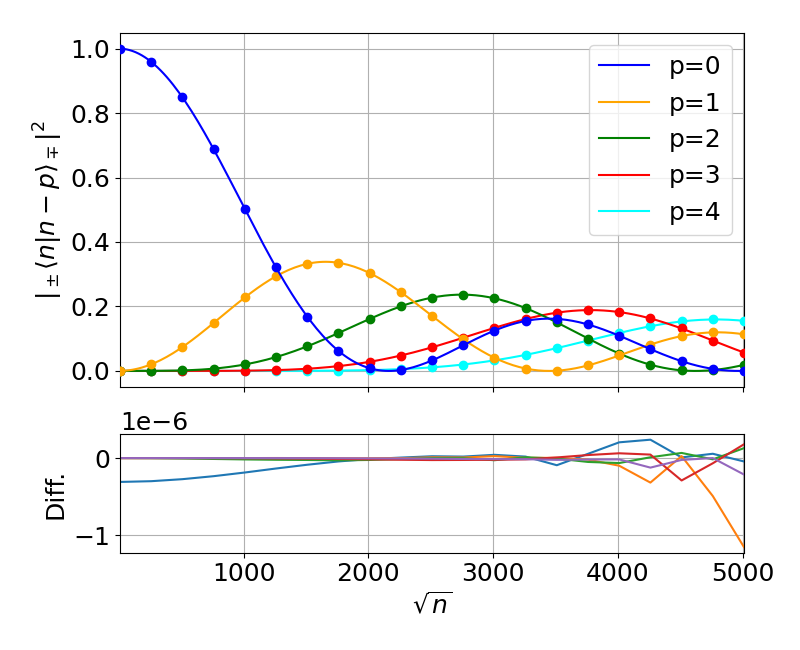}
    \caption{
    Comparison of the squared modules of the matrix elements for different $p$ values as function of square root of number of photons. Dots present full solutions based on Eq.~\eqref{eq:ln_overlap}, while solid lines shows approximated solutions based on Eq.~\eqref{eq:ln_overlap_semi}. Lower plot shows the difference between them.
    Used parameters: $\Omega_a/\omega_0 = 0.001$, and $\omega_L/\omega_0 = 0.9$.
    }
    \label{fig:comparisonMatrix}
\end{figure}
We now find the total spontaneous transition between TLS states in the semiclassical limit. For this purpose, we consider a coherent state $\ket{\alpha}$ with a large mean photon number $\bar{n} = |\alpha|^2 \gg 1$ as a reasonable approximation of the classical drive. Thus, the initial state is 
$\ket{\Psi_{i, \bar{n}}} \equiv \ket{i} \otimes \ket{\alpha} \otimes \ket{\text{vac}}$.
The coherent state in the Fock basis reads 
$\ket{\alpha} = \sum_n c_n \ket{n}$, with $c_n=(\alpha^n/\sqrt{n!})\exp(-|\alpha|^2/2) $. Acting with the $D(\mp\tfrac{\Omega_a}{4\omega_L})$ operator on both sides, we find the coherent state in the displaced Fock basis 
$D(\mp\tfrac{\Omega_a}{4\omega_L}) \ket{\alpha} = \ket{\alpha\mp\tfrac{\Omega_a}{4\omega_L}} = \sum_n c_n \ket{n}_\pm$. 
In the classical limit, the coupling strength is small 
$|\tfrac{\Omega_a}{4\omega_L}| \ll |\alpha|$, thus, we can approximate $\ket{\alpha\mp\tfrac{\Omega_a}{4\omega_L}}\approx\ket{\alpha}$ stating that it is also a coherent state with the same coefficients in the displaced Fock basis of eigenstates of the $H_0$ Hamiltonian.
As a consequence, the emission no longer originates from a single eigenstate but from a superposition thereof. However, since the system occupies high levels of the energy ladders, only states close to the initial one have significant overlap. We therefore denote the relevant matrix elements as ${}_\pm\braket{n}{n-p}_\mp$, where $|p|\ll n$ is the number of laser photons involved in the process. Under these conditions, and following Eq.~\eqref{eq:partial_gamma}, the total transition rate can be expressed as a sum of partial contributions, labeled by $p$:
\begin{equation}
    \Gamma_{i\bar{n} p} / \Gamma_0 
    \simeq 
    \left|{}_{s_i}\bigl\langle \left[\bar{n}\right] \bigl| \left[|\bar{n}\right]-p \bigr\rangle_{-s_i} \right|^2 \left(s_i1 + p\tfrac{\omega_L}{\omega_0}\right)^3 ,
\label{eq:partial_gamma_semi}
\end{equation}
with $[x]$ denoting the integer closest to $x$ (see Appendix~\ref{apx:totalrate}).

Eq.~\eqref{eq:ln_overlap} describes the matrix elements in the regime of large $n$ and small $p$. However, its direct evaluation may be numerically unstable due to the appearance of factorials. Thus, we derive the following approximated form
\begin{equation}
    {}_\pm\braket{n}{n-p}_\mp \approx (\pm)^p \ee^{\ii p \phi} J_p\left( \tfrac{|\Omega_a|}{\omega_L} \sqrt{n} \right),
\label{eq:ln_overlap_semi}
\end{equation}
where $J_p$ is the Bessel function of the first kind. The derivation is given in Appendix~\ref{apx:semiclassical}. The argument of the Bessel function is proportional to the classical electric field's amplitude, since $E\propto\sqrt{n}$. The comparison between the exact [based on Eq.~\eqref{eq:ln_overlap}] and approximated [based on Eq.~\eqref{eq:ln_overlap_semi}] squared modulus $|{}_\pm\braket{n}{n-p}_\mp|^2$ is presented in Fig.~\ref{fig:comparisonMatrix}, and shows very good agreement already for relatively small $n$ (relative difference $10^{-6}$). In quantum-optical experiments, typical laser beams contain between $10^6$ and $10^{20}$ photons per coherence time. Even for very weak beams the photon number is well above $10^4$, so coherent-state fluctuations are negligible and the dynamics is accurately described by Eq.~\eqref{eq:ln_overlap_semi}.
\begin{figure}[!t]
    \centering
    \includegraphics[width=0.99\linewidth]{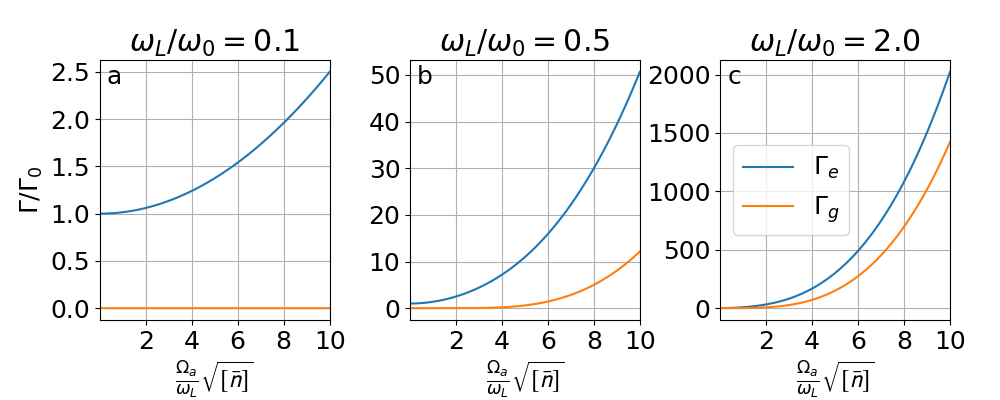}
    \caption{Rates for spontaneous emission $\Gamma_e$ (blue) and spontaneous absorption $\Gamma_g$ (orange) for different field frequencies in the semiclassical regime. Note, that the number of photons is different in each case. Coupling strength was set to $\Omega_a/\omega_0 = 0.001$.}
    \label{fig:approxVStotalGamma}
\end{figure}

According to Eq.~\eqref{eq:nRestriction}, a lower restriction for the $p$ values $p\geq - s_i\tfrac{\omega_0}{\omega_L}$ holds for ground and excited TLS initial state. Based on Eq.~\eqref{eq:partial_gamma_semi}, and on the form of the matrix elements from Eq.~\eqref{eq:ln_overlap_semi}, we find the emission rates for both the excited and ground states of the TLS in the semiclassical limit ($\Gamma_e$, and $\Gamma_g$, respectively)
\begin{subequations}
\label{eq:gammas_semi}
\begin{align}
    &\Gamma_{e} / \Gamma_0 
    =
    \left(1 + \tfrac{3|\Omega_a|^2}{2\omega_0^2} [\bar{n}] \right) + \Gamma_g / \Gamma_0,
    \label{eq:gamma_e}
    \\
    &\Gamma_g / \Gamma_0 
    = 
    \sum_{p \geq \omega_0/\omega_L}^\infty J^2_p\left( \tfrac{|\Omega_a|}{\omega_L} \sqrt{[\bar{n}]}\right) \left(p \tfrac{\omega_L}{\omega_0} -1 \right)^3. 
    \label{eq:gamma_g}
\end{align}
\end{subequations}
This form explicitly shows that the total spontaneous emission rate simplifies to $\Gamma_0$ for excited state, and vanishes for ground state when the driving field is absent, and always increases in presence
of the field. 
This stands in contrast to the result of the approximated lowest-order analysis where the suppression of emission rate was found in this approach \cite{Gladysz2025}. 
Moreover, 
the laser field unlocks the otherwise highly improbable excitation of the system via absorption of a reservoir photon. 

Based on Eqs.~\eqref{eq:gammas_semi}, we can distinguish two interesting regimes depending on the laser's frequency.
In the case of $\omega_L\ll\omega_0$, the term in Eq.~\eqref{eq:gamma_g} becomes negligible. 
In these conditions, the excitation from the ground state is suppressed, while spontaneous emission rate given by Eq.~\eqref{eq:gamma_e} becomes independent of the laser frequency and scales linearly with the average photon number, and hence quadratically with the field amplitude (see Fig.~\ref{fig:approxVStotalGamma}a).
In the regime of $\tfrac{\omega_L}{\omega_0}\gtrsim0.5$, the term $\Gamma_g / \Gamma_0$ in Eq.~\eqref{eq:gamma_g} becomes significant for small values of $p$. In that case, the probability of spontaneous transition from ground to excited state is relatively large, along with the increased spontaneous emission rate (see  Fig.~\ref{fig:approxVStotalGamma}b). The process efficiency increases with the blue detuning of the laser (Fig.~\ref{fig:approxVStotalGamma}c), since all the terms $p\geq 1$ contribute to the summation in Eq.~\eqref{eq:gamma_g}.

\section{Candidates for experimental realization}
This section comments on state-of-the-art experimental platforms that could potentially be exploited to observe the effects described in the previous sections. We include examples of experiments reaching the strong- and ultrastrong-coupling regimes for electronic and vibrational transitions in molecules, which have so far focused on the transverse coupling mechanism. Longitudinal coupling has been studied more extensively in superconducting circuits, while realizations via optical-lattice analogue physics are also feasible.

We first consider molecular and solid-state systems without inversion symmetry, which are natural candidates because their eigenstates can possess different permanent dipole moments. Relevant examples include polar molecules and molecular ions, semiconductor nanocrystals, color centers, and Rydberg molecules. Permanent dipole moments can range from several Debye in heteronuclear molecules \cite{liu2023} to tens or hundreds of Debye in colloidal nanorods and nanoplatelets \cite{li2003}, and to the kilodebye scale in trilobite Rydberg molecules \cite{booth2015}. When such large dipole moments are combined with strongly confined electromagnetic modes, appreciable longitudinal coupling becomes experimentally plausible. Moreover, if the permanent-dipole difference and the transition dipole are not collinear, the field polarization can be chosen to enhance the longitudinal coupling relative to the transverse one. Such a noncollinear arrangement occurs in a broad class of low-symmetry molecules and therefore does not represent an exceptional molecular configuration.

Ultrastrong coupling has been realized for vibrational transitions in polar molecular ensembles. Fukushima et al. \cite{fukushima2021} investigated vibrational modes with characteristic frequencies near 100~THz and reported collective coupling strengths reaching tens of terahertz. Although this experiment does not directly implement the single-emitter longitudinal Hamiltonian considered in our work, it demonstrates that normalized coupling strengths of order 0.1 can be achieved in polar molecular systems.

\begin{table*}[ht]
\centering
\begin{tabular}{|c|c|c|c|c|}
\hline 
\textbf{System,} &
\textbf{Qubit frequency} &
\textbf{Photon frequency} &
\textbf{Coupling strength} &
\textbf{Loss rate}
\\
\textbf{Reference} &
\textbf{$\omega_0/2\pi$} &
\textbf{$\omega_L/2\pi$} &
\textbf{$\Omega_a/2\pi$} &
\textbf{$\Gamma_0/2\pi$} \\
\hline

optical lattices &
$4.19$ kHz &
$4.32$ kHz &
trans.:
$0.587$ kHz &
estimated as tens of Hz
\\

Schweizer et al. \cite{schweizer2019}&
\ &
\ &
long.:
$7.95$ kHz &
\
\\
\hline

superconducting circuit &
$7.61$ GHz &
$4.35$ -- 
$6.10$ GHz &
$11.9$ MHz
(longitudinal) &
$<1$ MHz
\\

Potts et al. \cite{potts2025} &
\ &
\ &
\ &
\
\\
\hline

superconducting circuit &
$1.31$ GHz &
$5.15$ GHz &
$3.33$ GHz
(longitudinal) &
small
\\

Tomonaga et al. \cite{tomonaga2025} &
\ &
\ &
\ &
\
\\
\hline

superconducting circuit &
$1.27$ GHz &
$5.15$ GHz &
$3.45$ GHz
(longitudinal) &
small 
\\
Tomonaga et al. \cite{tomonaga2025} &
\ &
\ &
\ &
\
\\
\hline
\end{tabular}
\caption{Parameters of longitudinal-coupling experiments in selected analogue platforms.}\label{tab:tab}
\end{table*}

Large coupling strengths are also possible at the single-molecule level. Chikkaraddy et al. \cite{chikkaraddy2016} demonstrated coupling in the terahertz regime between a single molecule and a plasmonic nanocavity. The experiment was formulated in terms of an electronic transition and therefore involved predominantly transverse light-matter coupling. Nevertheless, it shows that the vacuum electric fields available in strongly confined plasmonic cavities are sufficiently large to produce very strong coupling to an individual molecular emitter. For a molecule whose two relevant states have substantially different permanent dipole moments, and with a suitable orientation of the cavity field, the same field-confinement strategy could, in principle, be used to enhance the longitudinal coupling.

The most direct realizations of the specific longitudinal interaction considered in our model have so far been achieved in engineered quantum systems. As a concrete example, we consider the superconducting-circuit experiment of Potts et al. \cite{potts2025}. In that experiment, a transmon with a characteristic transition frequency of approximately $\omega_0/2\pi \approx 7.61$~GHz was longitudinally coupled to a microwave resonator tunable over approximately $\omega_L/2\pi \approx 4.35–6.10$~GHz. The measured intrinsic longitudinal coupling strength was approximately $\Omega_a/2\pi \approx 11.9$~MHz, while the relevant resonator and qubit linewidths were below the megahertz scale.

An even larger normalized coupling has been reported in flux-qubit circuits. Tomonaga et al. \cite{tomonaga2025} studied flux qubits with characteristic frequencies of approximately 1.3 GHz coupled to a 5.15~GHz resonator. The reported coupling strengths were approximately 3.3-3.5~GHz. The ratio $\Omega_a/\omega_L \approx 0.65$ places the system deeply in the ultrastrong-coupling regime, where the effects considered in our manuscript are expected to be particularly pronounced.

A complementary realization of an effective ultrastrong longitudinal coupling has been implemented in an optical-lattice quantum simulator by Schweizer et al. \cite{schweizer2019}. The characteristic matter and mode frequencies were approximately 4.19~kHz and 4.32~kHz, respectively. The effective transverse and longitudinal coupling strengths were approximately 0.587~kHz and 7.95~kHz. The loss or decoherence scale can be estimated to be of the order of tens of Hertz, consistently with the observation of several coherent oscillation cycles. 

The summary of the experimental parameters achieved in different physical platforms is presented in Table~\ref{tab:tab}.

\section{Conclusions}
We have analyzed spontaneous emission in a driven polar TLS where the atom–laser interaction is governed by intrinsic dipole moments arising from broken inversion symmetry. Using a polaron transformation, we derived the dressed eigenstates and obtained analytical expressions for radiative transition rates.
We have shown that longitudinal coupling enables tunability of the overall radiative dynamics, which can lead to the suppression of spontaneous emission and the activation of excitation processes from the ground state. In the semiclassical limit, the dynamics is governed by multiphoton channels with Bessel-function weights.
These results identify driven polar quantum systems as a platform for engineering emission rates and spectral properties of light, with potential applications in tunable photon sources and quantum optical devices operating beyond the standard dipole-interaction picture.

An important direction for future research is a systematic treatment of the simultaneous transverse and longitudinal coupling of a TLS to the driving mode. Such an approach would allow to quantify how a finite transverse contribution may modify the transition rates derived here and whether interference between the longitudinal and transverse channels may play a role, e.g., enable additional resonances or modify selection rules. 
A complementary perspective concerns the extension of the present approach to multilevel or multiparticle systems. Finally, it would be interesting to characterize dynamics in the presence of a thermal reservoir, which would induce additional dissipative transitions, generally competing with the observed spontaneous processes and possibly reducing their significance, especially in systems with emission in the microwave domain.

\section*{Acknowledgments}
PG acknowledges support from ``Excellence Initiative -- Research University'' (IDUB) programme at Nicolaus Copernicus University in Toruń. 
KS acknowledges support from the National Science Centre, Poland, grant number 2023/50/E/ST3/00451. 
FVP acknowledges support from INFN through the project QUANTUM, and from PNRR MUR projects PRIN 20227JZKWP-``Emerging gauge theories: critical properties and quantum dynamics'', CN00000013-``Italian National Centre on HPC, Big Data and Quantum Computing'', and PE0000023-``National Quantum Science and Technology Institute'' (NQSTI).

\section*{Data availability}
The numerical calculations were performed using our open-source implementation~\cite{PGrepo}.

%\enlargethispage*{1cm}
\appendix
\makeatletter
\renewcommand{\section}{\@startsection{section}{1}{0pt}%
  {1.5ex plus 1ex minus .2ex}% space before
  {0.5ex }% space after
  {\normalfont\normalfont\bfseries}}
\makeatother

\section{Derivation of the TLS-laser Hamiltonian}\label{app:model}

The Hamiltonian \eqref{eq:H_0} originates from the addition of a free and an interaction term. The first contribution reads
\begin{equation}
    \tilde{H}_{\mathrm{free}} = \frac{1}{2} \hbar \tilde{\omega}_0 \sigma_z + \hbar\omega_L a^{\dagger}a ,
\end{equation}
where $\tilde{\omega}_0$ is the bare TLS transition frequency and $(a,a^{\dagger})$, satisfying the canonical commutation relation $[a,a^{\dagger}]=1$, are the bare laser mode operators. The interaction term is obtained starting from minimal coupling and after performing a Power-Zienau-Wolley transformation in the dipole approximation \cite{cohen2_1998,kibis2009,scala2020}, leading to
\begin{eqnarray}
    V = -\vec{d}\cdot\vec{E}_L(0) ,
\end{eqnarray}
where $d$ is the dipole operator and $\vec{E}_L(0)$ formally corresponds to the laser electric field evaluated in the TLS position $\vec{r}=0$, which reads
\begin{eqnarray}
    \vec{E}_L(0) = i \sqrt{\frac{\hbar\omega_L}{2\varepsilon_0}} \vec{\epsilon}(0) \left( \varphi(0) a - \varphi^*(0)a^{\dagger}\right)
\end{eqnarray}
where $\varphi(0)$ and $\vec{\epsilon}(0)$ are the amplitude and polarization of the mode, respectively. If we admit that both the bare TLS eigenstates $\ket{e}$ and $\ket{g}$ can have permanent dipole moments, but assume the laser coupling to the transition dipole to be negligible, e.g., due to polarization mismatch, the TLS-laser interaction Hamiltonian reads
\begin{equation}
    V = -i \sigma_z \otimes \tfrac{\hbar}{4} ( \Omega_a a^{\dagger} - \Omega^\star_a a) - i \openone \otimes ( \Omega_m a^{\dagger} - \Omega^\star_m a) ,
\end{equation}
with $\Omega_a$ defined as in Eq.~\eqref{eq:Omegaa}, and
\begin{equation}
    \Omega_m = - \frac{\vec{d}_{e} + \vec{d}_g}{2}\cdot\vec{\epsilon}(0) \sqrt{\frac{\omega_L}{2\hbar\varepsilon_0}} \varphi(0)^* .
\end{equation}
The sum $H_0=H_{\mathrm{free}}+V$ can be recast in the form \eqref{eq:H_0} upon defining a shifted mode operator
\begin{equation}
    b = ia + \frac{\Omega_m}{\omega_L}
\end{equation}
and a dressed excitation frequency
\begin{equation}
    \omega_0 = \tilde{\omega}_0 - \frac{\mathrm{Re}(\Omega_a^*\Omega_m)}{\omega_L} ,
\end{equation}
up to the irrelevant constant $-\hbar|\Omega_m|^2/\omega_L$.

\section{Polaron transformation properties}
\label{apx:polaron}
The algebra behind the transition from polaron transformation given by Eq.~\eqref{eq:polaron} to the final version with separated Hilbert spaces from Eq.~\eqref{eq:splitPolaron} reads
\begin{equation*}
    \begin{split}
    U 
    &= \sum_{n=0}^\infty 
    \tfrac{1}{n!} \left( \tfrac{1}{4\omega_L} \right)^n \left( \sigma_z \otimes (\Omega_a b^\dagger - \Omega_a^\star b) \right)^n 
    \nonumber
    \\
    &= \sum_{n=0}^\infty 
    \tfrac{1}{n!} \left( \tfrac{1}{4\omega_L} \right)^n \sigma_z^n \otimes (\Omega_a b^\dagger - \Omega_a^\star b)^n
    \\
    &\begin{split}
        = \sum_{n=0}^\infty 
        \tfrac{1}{n!} \left( \tfrac{1}{4\omega_L} \right)^n (\ket{e}\bra{e} + (-1)^n\ket{g}\bra{g}) \\ 
        \otimes (\Omega_a b^\dagger - \Omega_a^\star b)^n\nonumber
    \end{split}
    \\
    &= \ket{e}\bra{e} \otimes 
    \ee^{\tfrac{1}{4\omega_L} (\Omega_a b^\dagger-\Omega_a^\star b)}%}
    +\ket{g}\bra{g} \otimes 
    \ee^{\tfrac{-1}{4\omega_L} (\Omega_a b^\dagger - \Omega_a^\star b)}
    \nonumber
    \\
    &= \ket{e}\bra{e} \otimes D\left( \tfrac{\Omega_a}{4\omega_L} \right) 
    + \ket{g}\bra{g} \otimes D\left( -\tfrac{\Omega_a}{4\omega_L} \right). \nonumber
    \end{split}
\end{equation*}

Using $D(\alpha) D(\beta) = D(\alpha + \beta) \ee^{\ii \mathrm{Im}(\alpha \beta^\star)}$, we can show that system operators transform as $U \, \sigma_\pm \, U^\dagger=\sigma_\pm D\big( \pm \tfrac{\Omega_a}{2\omega_L} \big)$, while $ U \, b^{\cdot/\dagger} \, U^\dagger =  b^{\cdot/\dagger} - \sigma_z \tfrac{\Omega^{\cdot/\star}_a}{4\omega_L}$ is the general property of displacement operators.
The field's free evolution becomes 
\begin{equation}
    U \, \omega_L b^\dagger b \, U^\dagger = \omega_L b^\dagger b - \sigma_z \otimes \tfrac{1}{4} (\Omega_a b^\dagger + \Omega^\star_a b) + \tfrac{|\Omega_a|^2}{16\omega_L},
\label{apxeq:displaced_free}
\end{equation}
while the longitudinal coupling term reads
\begin{equation}
    U \sigma_z \otimes \tfrac{1}{4} (\Omega_a b^\dagger + \Omega_a^\star b) U^\dagger = 
    \sigma_z \otimes \tfrac{1}{4} (\Omega_a b^\dagger + \Omega^\star_a b) - \tfrac{|\Omega_a|^2}{8\omega_L}.
\label{apxeq:displaced_longitudinal}
\end{equation}
Summing Eqs.~\eqref{apxeq:displaced_free} and \eqref{apxeq:displaced_longitudinal}, we obtain the transformed Hamiltonian $\tilde{H}_0$ from Eq.~\eqref{eq:tildeH_0}. On the other hand, the transformation $U \sigma_x U^\dagger = U \sigma_+ U^\dagger + U \sigma_- U^\dagger$ provides form of the TLS-reservoir Hamiltonian $\tilde{H}_{IR}$ from Eq.~\eqref{eq:tildeH_IR}.

\section{Eigenstates}
\label{apx:eigenstates}

To find the overlap between the field states given in Eq.~\eqref{eq:ln_overlap}, we start with the definition of the laser's eigenstates given by Eq.~\eqref{eq:psi_in} noting that
\begin{equation}
    \begin{split}
    \ket{n}_\pm 
    &= D_\mp \ket{n} 
    = D_\mp \frac{1}{\sqrt{n!}} \left( b^\dagger \right)^n\ket{0} \\
    &= \frac{1}{\sqrt{n!}} \big( D_\mp b^\dagger D_\pm \big)^n \underbrace{D_\mp \ket{0}}_{\ket{0}_\pm} 
    \equiv \frac{1}{\sqrt{n!}} \left( d^\dagger_\pm \right)^n \ket{0}_\pm.
    \end{split}
\label{apxeq:n_pm}
\end{equation}
where $D_\pm \equiv D(\pm\tfrac{\Omega_a}{4\omega_L})$ and we used property $D_\pm D_\mp \equiv D_\pm D^\dagger_\pm = 1$. In result, we define the ladder of the displaced Fock states generated by the displaced mode operator $d_\pm = D_\mp b D_\pm$. Moreover, notice that the ground states are also eigenstates for the operators of the opposite sign
\begin{equation}
    \begin{split}
    d_\mp \ket{0}_\pm 
    &= D_\pm b D^2_\mp\ket{0} 
    = D_\pm b \ket{\mp2\frac{\Omega_a}{4\omega_L}} \\
    &= \mp \frac{\Omega_a}{2\omega_L} D_\pm \ket{\mp2\frac{\Omega_a}{4\omega_L}} 
    = \mp \frac{\Omega_a}{2\omega_L} \ket{0}_\pm,
    \end{split}
\label{apxeq:d_mp0_pm}
\end{equation}
where we used fact that for any coherent state $b\ket{\alpha} = \alpha \ket{\alpha}$. The above result along with the fact that for the ground states we have $d_\pm\ket{0}_\pm = 0$ provides immediately the following relation between the $d$ operators
\begin{equation}
    d_\pm = d_\mp \pm \frac{\Omega_a}{2\omega_L}.
\label{apxeq:d_pmd_mp}
\end{equation}

To evaluate ${}_\pm\braket{\ell}{n}_\mp$ we start by using Eq.~\eqref{apxeq:n_pm} on the bra state along with the definition from Eq.~\eqref{apxeq:d_pmd_mp}
\begin{equation*}
    \begin{split}
    &{}_\pm\braket{\ell}{n}_\mp 
    = {}_\pm\bra{0} \tfrac{1}{\sqrt{\ell !}}(d_\pm)^\ell \ket{n}_\mp\\
    &\quad = \tfrac{1}{\sqrt{\ell !}} {}_\pm\bra{0} \sum_{k=0}^{\mathrm{min}(\ell, n)} \tfrac{\ell!}{k! (l - k)!} \left( \pm\tfrac{\Omega_a}{2\omega_L} \right)^{\ell - k} 
    \underbrace{
        (d_\mp)^k \ket{n}_\mp
    }_{\sqrt{\tfrac{n!}{(n-k)!}} \ket{n-k}_\mp}
    \\
    &\quad = \sqrt{\ell! \, n!} \sum_{k=0}^{\mathrm{min}(\ell, n)} \tfrac{1}{k! (\ell - k)! \sqrt{(n - k)!}} \left( \pm\tfrac{\Omega_a}{2\omega_L} \right)^{\ell - k} \\
    &\qquad \times
    \underline{{}_\pm\bra{0} \left( d^\dagger_\mp \right)^{n - k}}
    \tfrac{1}{\sqrt{(n - k)!}} \ket{0}_\mp,
    \end{split}
\end{equation*}
next, applying the property \eqref{apxeq:d_mp0_pm} to the underlined term, we end up with
\begin{equation*}
    \begin{split}
    {}_\pm\braket{\ell}{n}_\mp 
    &= \sqrt{\ell! \, n!} \sum_{k=0}^{\mathrm{min}(\ell, n)} \tfrac{1}{k! (\ell - k)! (n - k)!} \left( \pm \tfrac{\Omega_a}{2\omega_L} \right)^\ell \left( \mp \tfrac{\Omega^\star_a}{2\omega_L} \right)^n \\
    & \quad \times (-1)^k \left| \tfrac{\Omega_a}{2\omega_L} \right|^{-2k} {}_\pm\braket{0}{0}_\mp,
    \end{split}
\end{equation*}
which, after rearrangement of the terms, is exactly the Eq.~\eqref{eq:ln_overlap} from the main text. Note that in the end, we have to use the overlap formula for the two coherent states ${}_\pm\braket{0}{0}_\mp = \exp\Big( {-\tfrac{1}{2}\left| \tfrac{\Omega_a}{2\omega_L} \right|^2} \Big)$.

For the case of same parity states overlapping, we can proceed similarly using properties of the harmonic oscillator
\begin{equation*}
    \begin{split}
    &{}_\pm\braket{\ell}{n}_\pm 
    = {}_\pm\bra{0} \tfrac{1}{\sqrt{\ell!}} d^\ell_\pm \ket{n}_\pm \\
    &\quad = \begin{cases}
        0, \; \text{if} \; \ell > n,
        \\
        \sqrt{\tfrac{n!}{\ell! \, (n-\ell)!}} \underline{{}_\pm\bra{0} \left( d^\dagger_\pm \right)^{n-\ell}} \tfrac{1}{\sqrt{(n-\ell)!}} \ket{0}_\pm, \; \text{if} \; \ell \leq n.
    \end{cases}
    \\
    &\quad = \delta_{\ell n},
    \end{split}
\end{equation*}
where the underlined term vanishes for $n\neq\ell$, which proves the statement from the main text.

\section{Resolvent theory}
\label{apx:resolvent}
To obtain the formula for the self energy given in Eq.~\eqref{eq:generalSelfEnergy} we start with the general form
\begin{equation}
    \Sigma_{i, n}(z)
    =
    \bra{\tilde{\Psi}_{i, n}} \tilde{H}_{IR} \sum_{n = 0}^{\infty} \left( \tfrac{1-\ket{\tilde{\Psi}_{i, n}} \bra{\tilde{\Psi}_{i, n}}}{z - \tilde{H}_0 - \tilde{H}_R} \tilde{H}_{IR} \right)^n \ket{\tilde{\Psi}_{i, n}},
\label{apxeq:generalSelfEnergy}
\end{equation}
where the state vector is given by Eq.~\eqref{eq:tildePsi_in}. Due to the form of the interaction Hamiltonian in Eq.~\eqref{eq:tildeH_IR}, we have $\bra{\tilde{\Psi}_{i, n}} \tilde{H}_{IR} \ket{\tilde{\Psi}_{i, n}} = 0$ as $\bra{\text{vac}} a_\epsilon^{\cdot/\dagger}(\vec{k}) \ket{\text{vac}} = 0$ for each vacuum mode. Additionally, we can express this self energy formula as a series of powers of coupling strength $g_\epsilon(\vec{k})$, and assuming weak coupling to the reservoir, truncate it on the linear term. In result Eq.~\eqref{apxeq:generalSelfEnergy} takes form of Eq.~\eqref{eq:generalSelfEnergy} provided in the main text with the correction $\mathcal{O} \left(g^2_\epsilon(\vec{k})\right)$.

Going further, the term $\bra{\tilde{\Psi}_{i,n}} \tilde{H}_{IR}$ (and its hermitian conjugation) appearing in Eq.~\eqref{eq:generalSelfEnergy} can be written explicitly as
\begin{widetext}
\begin{equation}
    \begin{split}
    \bra{\tilde{\Psi}_{i,n}} \tilde{H}_{IR} 
    &= \int \dd \vec{k} \sum_\epsilon \Bigg( g^\star_\epsilon(\vec{k})
    \underbrace{
        \bra{\tilde{\Psi}_{i,n}} D\left( \tfrac{\Omega_a}{2\omega_L} \right) \sigma_+a_\epsilon(\vec{k})
    }_{\delta_{ie} \bra{g} \otimes \bra{n} D\left( \tfrac{\Omega_a}{2\omega_L} \right) \otimes \bra{1_\epsilon(\vec{k})}}
    + g_\epsilon(\vec{k}) \underbrace{
        \bra{\tilde{\Psi}_{i,n}} D\left( -\tfrac{\Omega_a}{2\omega_L} \right) \sigma_-a^\dagger_\epsilon(\vec{k})
    }_{\bra{\mathrm{vac}} a^\dagger_\epsilon(\vec{k})=0} \Bigg)
    \\
    &+ \int \dd \vec{k} \sum_\epsilon \Bigg( g_\epsilon(\vec{k})
    \underbrace{
        \bra{\tilde{\Psi}_{i,n}} D\left( \tfrac{\Omega_a}{2\omega_L} \right)\sigma_+ a^\dagger_\epsilon(\vec{k})
    }_{\bra{\mathrm{vac}} a^\dagger_\epsilon(\vec{k})=0}
    + g^\star_\epsilon(\vec{k}) \underbrace{
        \bra{\tilde{\Psi}_{i,n}} D\left( -\tfrac{\Omega_a}{2\omega_L} \right) \sigma_- a_\epsilon(\vec{k})
    }_{\delta_{ig} \bra{e} \otimes \bra{n} D\left( -\tfrac{\Omega_a}{2\omega_L} \right) \otimes \bra{1_\epsilon(\vec{k})}} \Bigg),
    \end{split}
\label{apxeq:parts}
\end{equation}
\end{widetext}
where first integral corresponds to the spontaneous emission (rotating terms), while the second one to the less probable yet non-negligible in this case spontaneous absorption (counter-rotating terms). Effectively, for given $i$ only one element in both brackets survives. Note that $\delta_{ie}$ appears in the spontaneous emission terms while $\delta_{ig}$ in the spontaneous absorption ones providing restrictions for the evolution of the TLS, what is discussed in the main text.

Next, the spectral decomposition of the middle operator from Eq.~\eqref{eq:generalSelfEnergy} reads
\begin{equation}
    \begin{split}
    &\tfrac{1}{z - \tilde{H}_0 - \tilde{H}_R} 
    = \\
    &
    \sum_{i \in \{e, g\}} \sum_{n,m = 0}^\infty \sum_{m = 0}^\infty \sum_{\epsilon\in\{1, 2\}} 
    \int \dd \vec{k} \tfrac{\ket{i}\bra{i} \otimes \ket{n}\bra{n} \otimes \ket{m_\epsilon(\vec{k})}\bra{m_\epsilon(\vec{k})}}{z - s_i\tfrac{\omega_0}{2} - n \omega_L -  m_\epsilon(\vec{k})\omega(\vec{k})}.
    \end{split}
    \label{apxeq:resolventSpectral}
\end{equation}
Thus, inserting Eqs.~\eqref{apxeq:parts} and~\eqref{apxeq:resolventSpectral} into Eq.~\eqref{eq:generalSelfEnergy} we obtain the explicit formula for the approximated self energy 
\begin{widetext}
\begin{equation}
    \begin{split}
    \Sigma_{i,n}(z) 
    =& \int \dd\vec{k} \sum_{\epsilon} \int \dd \vec{k^\prime} \sum_{\epsilon^\prime} \sum_{n^\prime,m} \int \dd \vec{k}^{\prime\prime} \sum_{\epsilon^{\prime\prime}} g^\star_\epsilon(\vec{k}) g_{\epsilon^{\prime\prime}}(\vec{k}^{\prime\prime}) 
    \tfrac{\left| \bra{n} D\left( s_i \tfrac{\Omega_a}{2\omega_L} \right) \ket{n^\prime} \right|^2\braket{1_{\epsilon}(\vec{k})}{m_{\epsilon^\prime}(\vec{k^\prime})} \braket{m_{\epsilon^\prime}(\vec{k^\prime})}{1_{\epsilon^{\prime\prime}}(\vec{k}^{\prime\prime})}}{z + s_i\tfrac{\omega_0}{2} - n^\prime \omega_L - m_{\epsilon^\prime}(\vec{k}^\prime) \omega(\vec{k}^\prime)}
    \\
    =& \int \dd\vec{k} \sum_{\epsilon} \int \dd \vec{k^\prime} \sum_{\epsilon^\prime} \sum_{n^\prime, m} \int \dd \vec{k}^{\prime\prime} \sum_{\epsilon^{\prime\prime}}
    g^\star_\epsilon(\vec{k}) g_{\epsilon^{\prime\prime}}(\vec{k}^{\prime\prime}) 
    \underbrace{
        \braket{1_{\epsilon^\prime}(\vec{k})}{m_{\epsilon^{\prime}}(\vec{k}^{\prime})} \braket {m_{\epsilon^{\prime}}(\vec{k}^{\prime})}{1_{\epsilon^{\prime\prime}}(\vec{k}^{\prime\prime})}
    }_{\delta_{1m}\delta_{\epsilon\epsilon^{\prime}}\delta(\vec{k}-\vec{k}^{\prime})\delta_{\epsilon^\prime\epsilon^{\prime\prime}}\delta(\vec{k}^\prime-\vec{k}^{\prime\prime})} 
    \tfrac{\left| \bra{n} D\left( s_i \tfrac{\Omega_a}{2\omega_L} \right) \ket{n^\prime} \right|^2}{z + s_i\tfrac{\omega_0}{2} - n^\prime \omega_L - m_{\epsilon^\prime}(\vec{k}^\prime) \omega(\vec{k}^\prime)} 
    \\
    =& \sum_{n^\prime} \left| {}_{s_i}\braket{n}{n^\prime}_{-s_i} \right|^2 \int \dd \vec{k} \sum_\epsilon \tfrac{|g_{\epsilon}(\vec{k})|^2}{z + s_i\tfrac{\omega_0}{2} - n^\prime\omega_L - \omega(\vec{k})}.
    \end{split}
\label{apxeq:fullSelfEnergy}
\end{equation}
\end{widetext}
Notice the change of sign of the TLS energy in the denominator in comparison to Eq.~\eqref{apxeq:resolventSpectral} as we had flip of the states by the $\sigma_\pm$ operators from Eqs.~\eqref{apxeq:parts}.

To calculate the self energy for the given initial state $\ket{\Psi_{i,n}}$ we start with the form Eq.~\eqref{apxeq:fullSelfEnergy} inserting state's energy along with infinitely small imaginary part $\ii 0$
\begin{equation}
    \begin{split}
    &\Sigma_{i, n} \left( s_i\tfrac{\omega_0}{2} + n \omega_L + \ii0 \right) 
    \\
    &= \sum_{n^\prime}  \left| {}_{s_i}\braket{n}{n^\prime}_{-s_i} \right|^2 \int \dd \vec{k} \sum_\epsilon \tfrac{|g_{\epsilon}(\vec{k})|^2}{s_i\omega_0 + (n - n^\prime) \omega_L - \omega(\vec{k}) + \ii0}
    \\
    &= \sum_{n^\prime}  \left| {}_{s_i}\braket{n}{n^\prime}_{-s_i} \right|^2 
    \Big[ \mathrm{P.V.} \int \dd \vec{k} \sum_\epsilon \tfrac{|g_{\epsilon}(\vec{k})|^2}{s_i \omega_0 + (n - n^\prime)\omega_L - \omega(\vec{k})} 
    \\
    & \quad - \ii \pi \int \dd \vec{k} \sum_\epsilon |g_{\epsilon}(\vec{k})|^2 \delta \left( s_i \omega_0 + (n - n^\prime) \omega_L - \omega(\vec{k}) \right) \Big]
    \\
    &= \sum_{n^\prime} \left( \Delta_{inn^\prime} - \tfrac{\ii}{2} \Gamma_{inn^\prime} \right),
    \end{split}
\label{apxeq:selfEnergy}
\end{equation}
where P.V. stands for the principal value while $\delta$ is Dirac delta distribution. With that we end up with the form from Eq.~\eqref{eq:selfEnergy}. From the Eq.~\eqref{apxeq:selfEnergy} we can write exact forms of the quantities $\Delta_{inn^\prime}$ and $\Gamma_{inn^\prime}$:
\begin{subequations}\label{apxeq:deltaGamma}
    \begin{align}
    \Delta_{inn^\prime} 
    &=  \left| {}_{s_i}\braket{n}{n^\prime}_{-s_i} \right|^2  \mathrm{P.V.} \int \dd \vec{k} \sum_\epsilon \tfrac{|g_{\epsilon}(\vec{k})|^2}{s_i\omega_0 + (n - n^\prime)\omega_L - \omega(\vec{k})},
    \label{apxeq:delta}
    \\
    \begin{split}
    \Gamma_{inn^\prime} 
    &=  \left| {}_{s_i}\braket{n}{n^\prime}_{-s_i} \right|^2  
    2\pi \int \dd \vec{k} \sum_\epsilon |g_{\epsilon}(\vec{k})|^2 
    \\
    & \quad \times \delta \left( s_i\omega_0 + (n - n^\prime) \omega_L - \omega(\vec{k}) \right).
    \end{split}
    \label{apxeq:gamma}
    \end{align}
\end{subequations}
Form of the $\Gamma$ rates from Eq.~\eqref{apxeq:gamma} gives us the restriction for the allowed transitions as $\omega(\vec{k})$ has to be non-negative, hence the argument of the Dirac delta
\begin{equation*}
    s_i \omega_0 + (n - n^\prime) \omega_L - \omega(\vec{k}) \geq0,
\end{equation*}
which provides restrictions for $n^\prime$ numbers as given in the main text by Eq.~\eqref{eq:nRestriction}. This, in consequence, determine the possible transitions between the energy levels to be only downwards and between the energy ladders.

To compute the gamma rates we have to evaluate the integral, and for that we also need to apply proper dispersion relation. In this paper we use 3D case, hence the system-reservoir coupling strength can be written as 
\begin{equation}
    g_\epsilon(\vec{k}) 
    = \sqrt{\tfrac{\omega(\vec{k})}{2\hbar \epsilon_0 (2\pi)^3}} \vec{d} \cdot \vec{\ee}_\epsilon(\vec{k}) 
    \Rightarrow 
    \sum_{\epsilon} \left| g_\epsilon(\vec{k}) \right|^2 
    = \tfrac{\omega(\vec{k}) |\vec{d}|^2}{3\hbar \epsilon_0 (2\pi)^3},
\end{equation}
where $\vec{d}$ is the transition electric dipole, and $\vec{\ee}_\epsilon(\vec{k})$ is the unit vector along given field mode. The averaging over all polarization directions in the sum provided $2/3$ term ($1/3$ for each polarization). The dispersion relation in this case is $\omega(\vec{k}) = c|\vec{k}|$, while the transition from Cartesian to spherical coordinates is $\dd \vec{k} = k^2\sin\phi \, \dd k \,\dd \theta \, \dd\phi$ with proper boundaries, hence from Eq.~\eqref{apxeq:gamma} the gamma rates read
\begin{equation}
    \begin{split}
    \Gamma_{inn^\prime} 
    =& \left| {}_{s_i}\braket{n}{n^\prime}_{-s_i} \right|^2
    2 \pi \int_0^\infty \dd k 
    \underbrace{
        \int_0^{\pi} \dd \phi \sin \phi \int_0^{2\pi} \dd \theta}_{4\pi}
        \\
        & \quad \times k^2 \tfrac{ck|\vec{d}|^2}{3 \hbar \epsilon_0 (2\pi)^3} \delta(s_i\omega_0 + (n - n^\prime) \omega_L - ck)
    \\
    =& \left| {}_{s_i}\braket{n}{n^\prime}_{-s_i} \right|^2 \tfrac{(s_i \omega_0 + (n - n^\prime) \omega_L)^3 |\vec{d}|^2}{3 \hbar \epsilon_0 \pi c^3}.
    \end{split}
\label{apxeq:3DGammainn}
\end{equation}

In the absence of the longitudinal coupling, for TLS prepared in the excited state, and coupled to the 0K reservoir, the spontaneous emission is described by the exponential decay with rate \begin{equation}
    \Gamma_0 
    = \tfrac{\omega^3_0 |\vec{d}|^2}{3 \hbar \epsilon_0 \pi c^3}. 
\label{apxeq:gamma3D0}
\end{equation}
We can use this quantity as a reference for the investigation of the gamma rates in case of longitudinal coupling by rewriting the 3D free space expression given in Eq.~\eqref{apxeq:3DGammainn}, which provides us the form given in the main text in Eq.~\eqref{eq:partial_gamma}.

\section{Atomic transition rates with an arbitrary state of the laser}\label{apx:totalrate}
In this section, we clarify the definition of the quantity of Eq.~\eqref{eq:partial_gamma_semi} and the related approximations. Consider an initial state in the form
\begin{equation}
    \ket{\Psi_0^{i}} = \ket{i}\otimes\ket{\psi_L}\otimes\ket{\mathrm{vac}},
\label{apxeq:initial_state}
\end{equation}
with $i\in\{e,g\}$ and
\begin{equation}
    \ket{\psi_L} = \sum_{n=0}^{\infty} c_n^{i} \ket{n}_{s_i}
\label{apxeq:initial_laser}
\end{equation}
an arbitrary state of the laser, expanded for convenience on the basis of associated to $\ket{i}$ in the atom-laser eigenstates. We can define the total rate of decay from the atomic level $\ket{i}$ as
\begin{equation}
    \Gamma_i = - \frac{d}{dt} \ln P_i(t) \Bigl|_{t=0},
\label{apxeq:gamma_def}
\end{equation}
where
\begin{equation}
    P_i(t) = \mathrm{Tr} \left[ \ket{i}\bra{i} U(t)\ket{\Psi_0^{i}} \bra{\Psi_0^{i}} U^{\dagger}(t)  \right],
\label{apxeq:trace_general}
\end{equation}
with $U(t)$ the global evolution operator. Using the basis $\{\ket{i}\otimes\ket{n}_{s_i}\}$ for the atom-laser space, computation of the trace from~\eqref{apxeq:trace_general} with explicit laser state from~\eqref{apxeq:initial_laser} yields
\begin{equation}
    P_i(t) = \sum_{n=0}^{\infty} |c_n^{i}|^2 P_{i,n}(t) ,
\label{apxeq:P_i}
\end{equation}
where
\begin{equation}
    P_{i,n}(t) = \sum_R \bra{i} \otimes {}_{s_i}\bra{n} \otimes \bra{R} U(t) \ket{i} \otimes \ket{n}_{s_i} \otimes \ket{R} ,
\end{equation}
with the sum running on any basis of the radiation space, the survival probability of the $\{\ket{i}\otimes\ket{n}_{s_i}\}$ atom-laser eigenstate, approximately reading 
\begin{equation}
    P_{i,n}(t) \simeq \exp \left( - \Gamma_{i,n} t \right) = \exp\left( - t \sum_{n'} \Gamma_{inn'}  \right) .
\label{apxeq:P_in}
\end{equation}
Hence, based on Eq.~\eqref{apxeq:gamma_def} for $P_i(t)$ given by Eq.~\eqref{apxeq:P_i} with terms from Eq.~\eqref{apxeq:P_in} we have
\begin{equation}
    \Gamma_e = \sum_{n=0}^{\infty} |c_n^{i}|^2 \Gamma_{i,n} .
\end{equation}
If the photon distribution is peaked around $\bar{n}$, with spread $\Delta n \ll \bar{n}$, the above quantity can be safely approximated as 
\begin{equation}
    \Gamma_e \simeq \Gamma_{i,[\bar{n}]} = \sum_{p} \Gamma_{i\bar{n}p} ,
\end{equation}
where $[x]$ denotes the closest integer to $x$ and $\Gamma_{i\bar{n}p}:=\Gamma_{i[\bar{n}][\bar{n}]-p}$. In the Markovian approximation, assuming that spontaneous emission does not significantly alter the initial photon distribution, the computed emission rates are valid for the whole duration of the experiment.

\section{Semiclassical limit}
\label{apx:semiclassical}
Here, we provide derivation of the approximated expression for ${}_\pm\braket{n}{n-p}_\mp$ for large $n$ and small $p$ so $n\gg |p|$ ($p$ can be negative). Let us start with the explicit form from Eq.~\eqref{eq:ln_overlap} for $\ell = n$, $n=n-p$
\begin{equation}
    \begin{split}
    {}_\pm\braket{n}{n-p}_\mp 
    =& (\pm)^n (\mp)^{n-p} \sqrt{n! \, (n-p)!} \; \ee^{-\tfrac{1}{2}\left| \tfrac{\Omega_a}{2\omega_L} \right|^2} \ee^{\ii p\phi } \times
    \\
    &\underline{\sum_{k=0}^{\mathrm{min}(n, n-p)} \tfrac{(-1)^k}{k! (n - k)! (n - p - k)!} \left| \tfrac{\Omega_a}{2\omega_L} \right|^{2n -p -2k}}
    \end{split}
\label{apxeq:overlapnp}
\end{equation}
First, we take care of the sum underlined above, renaming the summation index $k \rightarrow n-k$ we have
\begin{equation}
    \begin{split}
    &\sum_{k=\mathrm{max(0, p)}}^n \tfrac{(-1)^{n - k}}{(n - k)! \, k! \, (k - p!)} \left| \tfrac{\Omega_a}{2\omega_L} \right|^{2k - p}
    \\
    &= \tfrac{(-1)^n}{(n-p)!} \left| \tfrac{\Omega_a}{2\omega_L} \right|^{-p} \sum_{k=\mathrm{max(0, p)}}^n \tfrac{(n-p)!}{(n-k)!(k-p)!}\tfrac{(-1)^k}{k!} \left| \tfrac{\Omega_a}{2\omega_L} \right|^{2k}
    \\
    &\approx \tfrac{(-1)^n}{(n-p)!} \left| \tfrac{\Omega_a}{2\omega_L} \right|^{-p} L^{(-p)}_n\left( \left| \tfrac{\Omega_a}{2\omega_L} \right|^2 \right),
    \end{split}
\label{apxeq:overlapLaguerre}
\end{equation}
where we approximated $\mathrm{max}(0,p) \approx 0$ since the summation goes up to the high values of $n$, and immediately recognized the sum as an associated Laguerre polynomials $L^{(-p)}_n(x)$. The asymptotic behavior for such polynomials for fixed argument and large $n$ are given by Plancheler-Rotach formula
\begin{equation}
    \begin{split}
    L_n^{(-p)}\left( \left| \tfrac{\Omega_a}{2\omega_L} \right|^2 \right)
    = 
    n^{-p/2} \left| \tfrac{\Omega_a}{2\omega_L} \right|^{p} \ee^{\tfrac{1}{2}\left| \tfrac{\Omega_a}{2\omega_L} \right|^2} J_{-p}\left( 2 \tfrac{|\Omega_a|}{2\omega_L} \sqrt{n} \right) 
    \\
    \times \left[ 1 + O\left( n^{-1/2} \right) \right].
    \end{split}
\label{apxeq:Plancheler}
\end{equation}
For the more detailed derivation see Appendix \ref{apx:plancheler}. Dropping the $O(n^{-1/2})$ term and putting Eqs.~\eqref{apxeq:overlapLaguerre}, and~\eqref{apxeq:Plancheler} into Eq.~\eqref{apxeq:overlapnp} we have approximated form
\begin{equation}
    {}_\pm\braket{n}{n-p}_\mp \approx (\mp)^p \sqrt{\tfrac{n!}{(n-p)!}} n^{-p/2} \, \ee^{\ii p \phi} \, J_{-p}\left( 2 \tfrac{|\Omega_a|}{2\omega_L} \sqrt{n} \right).
\label{apxeq:overlapnl_intermediate}
\end{equation}
The square root of factorials above can also be approximated for large $n$ as follows
\begin{equation}
    \begin{split}
    \sqrt{\tfrac{n!}{(n-p)!}} 
    &= 
    n^{p/2} \sqrt{1 \cdot \left( 1 - \tfrac{1}{n} \right) \cdot \left( 1 - \tfrac{2}{n} \right) \cdots \left( 1 - \tfrac{p-1}{n} \right)}
    \\
    &=
    n^{p/2} \left[ 1 + O\left( n^{-1/2} \right) \right].
    \end{split}
\label{apxeq:leftoversn}
\end{equation}
Dropping again terms of the order of $O(n^{-1/2})$, putting result from Eq.~\eqref{apxeq:leftoversn} into Eq.~\eqref{apxeq:overlapnl_intermediate}, and using Bessel function property $J_{-p}(x) = (-1)^p J_p(x)$, we end up with the formula Eq.~\eqref{eq:ln_overlap_semi} from the main text.

According to Eq.~\eqref{eq:nRestriction} we have lower restriction for the $p$ values $p\geq - s_i\tfrac{\omega_0}{\omega_L}$ for ground and excited TLS initial state. Based on Eq.~\eqref{eq:partial_gamma}, we write explicitly total $\Gamma_{i,n}$ rates as
\begin{equation}\label{apxeq:largenApprox}
    \begin{split}
    \Gamma_{i,n} / \Gamma_0
    &= 
    \sum_{p \geq -s_i\omega_0/\omega_L}^n \left| {}_{s_i}\braket{n}{n-p}_{-s_i} \right|^2 \left( s_i1 + p \tfrac{\omega_L}{\omega_0} \right)^3 
    \\
    &\approx 
    \sum_{p \geq -s_i\omega_0/\omega_L}^\infty J^2_p\left( \tfrac{|\Omega_a|}{\omega_L} \sqrt{n}\right) \left( s_i1 + p \tfrac{\omega_L}{\omega_0} \right)^3.
    \end{split}
\end{equation}
Formally, we truncate the summation at $p=n$ as the lowest possible state the system can end up in is $\ket{0}_\pm$, however, in our case we can extend it to $\infty$ for easier analysis of the properties. Note that for $\omega_L\geq\omega_0$ only nonnegative values of $p$ are available. 

In general, the formula Eq.~\eqref{apxeq:largenApprox} is not easy to analyze but several observations can be made with certain assumptions. Note that the sum can be rewritten as
\begin{equation}
    \begin{split}
    \Gamma_{i,n} / \Gamma_0 
    &=
    \sum_{p = -\infty}^\infty J^2_p\left( \tfrac{|\Omega_a|}{\omega_L} \sqrt{n}\right) \left( s_i1 + p\tfrac{\omega_L}{\omega_0} \right)^3 
    \\
    &+ \sum_{p > s_i\omega_0/\omega_L}^{\infty} J^2_p\left( \tfrac{|\Omega_a|}{\omega_L} \sqrt{n}\right) \left( -s_i1 + p \tfrac{\omega_L}{\omega_0} \right)^3.
    \end{split}
\label{apxeq:largenApproxRephrased}
\end{equation}
The first sum in Eq.~\eqref{apxeq:largenApproxRephrased} can be split into four sums according to the expansion of the bracket
\begin{equation*}
    \left( s_i1 + p\tfrac{\omega_L}{\omega_0} \right)^3 = s_i + 3p\tfrac{\omega_L}{\omega_0} + s_i 3p^2\left( \tfrac{\omega_L}{\omega_0}\right)^2 + p^3\left( \tfrac{\omega_L}{\omega_0}\right)^3,
\end{equation*}
hence, we have 
\begin{subequations}\label{apxeq:besselProperties}
    \begin{align}
    &s_i\sum_{p=-\infty}^\infty J^2_p\left( \tfrac{|\Omega_a|}{\omega_L} \sqrt{n}\right) 
    = 
    s_i1,
    \\
    &3 \tfrac{\omega_L}{\omega_0} \sum_{p=-\infty}^\infty p J^2_p\left( \tfrac{|\Omega_a|}{\omega_L} \sqrt{n}\right) 
    = 
    0,
    \\
    &s_i3 \left( \tfrac{\omega_L}{\omega_0} \right)^2 \sum_{p=-\infty}^\infty p^2 J^2_p\left( \tfrac{|\Omega_a|}{\omega_L} \sqrt{n}\right) 
    = 
    s_i3 \left( \tfrac{\omega_L}{\omega_0} \right)^2 \tfrac{|\Omega_a|^2}{2\omega_L^2}n,
    \\
    &\left( \tfrac{\omega_L}{\omega_0} \right)^3 \sum_{p=-\infty}^\infty p^3 J^2_p\left( \tfrac{|\Omega_a|}{\omega_L} \sqrt{n}\right) 
    = 
    0,
    \end{align}
\end{subequations}
where terms with odd powers of $p$ vanish simply because of the symmetry of the whole sum. The second sum from Eq.~\eqref{apxeq:largenApproxRephrased} if compared with Eq.~\eqref{apxeq:largenApprox} is easily recognized as $\Gamma_{j,n}$ rate for the opposite TLS state $j\neq i$, what in result provides us with the following form
\begin{equation}
    \Gamma_{i,n} / \Gamma_0
    = 
    s_i\left(1 + \tfrac{3|\Omega_a|^2}{2\omega_0^2} n \right)
    + \Gamma_{j,n}/\Gamma_0.
\label{apxeq:convolutedGammas}
\end{equation}
While this approach does not provide us with any exact formula we know that process in which TLS goes from ground to excited state is less effective, hence $\Gamma_{g,n}$ can be expressed exactly by Eq.~\eqref{apxeq:largenApprox}, and used as a correction to the $\Gamma_{e,n}$ given by Eq.~\eqref{apxeq:convolutedGammas}, hence
\begin{subequations}
\begin{align}
    &\Gamma_{e,n} / \Gamma_0 
    =
    \left(1 + \tfrac{3|\Omega_a|^2}{2\omega_0^2} n \right) + \Gamma_{g,n} / \Gamma_0.
    \\
    &\Gamma_{g,n} / \Gamma_0 
    = 
    \sum_{p \geq \omega_0/\omega_L}^\infty J^2_p\left( \tfrac{|\Omega_a|}{\omega_L} \sqrt{n}\right) \left(p \tfrac{\omega_L}{\omega_0} -1 \right)^3,
\end{align}
\end{subequations}
which is the formula given in Eq.~\eqref{eq:gammas_semi} if we rewrite $n\rightarrow[\bar{n}]$ referencing the mean number of photons from the coherent state $\ket{\alpha}$ and drop $n$ subscript in $\Gamma$ rates.

\section{Derivation of the Plancheler-Rotach formula}
\label{apx:plancheler}
To show validity of the Plancheler-Rotach formula given in Eq.~\eqref{apxeq:Plancheler}, we start with the equation 
\begin{equation}
    xy^{\prime\prime}(x) + (\alpha + 1 - x) y^\prime(x) + n y(x) 
    = 0.
\label{apxeq:start_eq}
\end{equation}
The exact solution for this are associated Laguerre polynomials $L^{(\alpha)}_n(x)$. To find the approximate solution for large $n$ and $x \ll n$ we use substitution
\begin{equation}
    y(x) 
    = x^{-\alpha/2} \ee^{x/2} u(x),
\label{apxeq:start_solution}
\end{equation}
hence we have
\begin{equation*}
    \begin{split}
    &y^\prime 
    =\left[ \left( \frac{1}{2} - \frac{\alpha}{2x} \right)u + u^\prime\right]x^{-\alpha/2}\ee^{x/2},
    \\
    &y^{\prime\prime} 
    = \left[ u^{\prime\prime} + \left( 1 - \frac{\alpha}{x} \right) u^\prime + \left( \frac{1}{4} - \frac{\alpha}{2x} + \frac{\alpha}{2x^2} + \frac{\alpha^2}{4x^2} \right) u \right]
    \\
    &\qquad\times x^{-\alpha/2}\ee^{x/2}.
    \end{split}
\end{equation*}
which, after division by $x$, leads us to the new form of Eq.~\eqref{apxeq:start_eq}
\begin{equation*}
    u^{\prime\prime} + \frac{1}{x} u^\prime + \left( -\frac{1}{4} + \frac{2n + \alpha + 1}{2x} - \frac{\alpha^2}{4x^2} \right) u = 0.
\end{equation*}
To proceed we change the variable $z = 2\sqrt{nx}$, and changing the derivatives accordingly
\begin{equation*}
    \begin{split}
    &\frac{\dd}{\dd x} = \frac{\dd z}{\dd x} \frac{\dd}{\dd z} 
    \equiv \sqrt{\frac{n}{x}} \frac{\dd}{\dd z},
    \\
    &\frac{\dd^2}{\dd x^2} = \frac{\dd}{\dd x} \left( \sqrt{\frac{n}{x}} \frac{\dd}{\dd z} \right) \equiv \frac{n}{x} \frac{\dd^2}{\dd z^2} -\frac{1}{2x} \sqrt{\frac{n}{x}} \frac{\dd}{\dd z},
    \end{split}
\end{equation*}
after multiplying each side by $4x^2$ we have
\begin{equation*}
    \begin{split}
    &\underbrace{4xn}_{z^2}
    \, u^{\prime\prime} + 
    \underbrace{2\sqrt{nx}}_{z} \, u^\prime 
    \\
    &+ 
    [\underbrace{4nx}_{z^2} + \underbrace{2x}_{z^2/2n}(\alpha + 1) - \underbrace{x^2}_{z^4/16n^2} - \alpha^2] \, u 
    = 0.
    \end{split}
\end{equation*}
For large $n$ leading terms are
\begin{equation*}
    z^2 u^{\prime\prime} + z u^\prime + (z^2 - \alpha^2) u = 0,
\end{equation*}
and the solution to this approximated equation are Bessel functions of the first kind $J_\alpha(z)$, more precisely
\begin{equation}
    u(z) = C J_\alpha(z),
\label{apxeq:bessel_solution}
\end{equation}
where $C$ is a constant that has to be evaluated for the behavior of the original equation near zero. Inserting approximated solution~\eqref{apxeq:bessel_solution} into Eq.~\eqref{apxeq:start_solution}, we found near zero expansion
\begin{equation}
    \begin{split}
    y(x) 
    &= 
    C \, \ee ^{x/2} \, x^{-\alpha/2} J_\alpha(2\sqrt{nx}) 
    \\
    &\approx
    C \, \ee ^{x/2} \, x^{-\alpha/2} \, \frac{1}{\Gamma(\alpha + 1)} \, (nx)^{\alpha/2} 
    \\
    &= 
    \frac{C}{\Gamma(\alpha + 1)} \, \ee^{x/2} \, n^{\alpha/2},
    \\
    y(0) 
    &= \frac{C}{\Gamma(\alpha + 1)}\, n^{\alpha/2}.
    \end{split}
    \label{apxeq:near_zero_y}
\end{equation}
On the other hand, we have exact solution to the initial Eq.~\eqref{apxeq:start_eq}, which are associated Laguerre polynomials, hence based on the Stirling's approximation we get
\begin{equation}
    L^{(\alpha)}_n(0) 
    = \frac{\Gamma(n + \alpha + 1)}{n! \, \Gamma(\alpha + 1)} 
    \approx \frac{n^{\alpha}}{\Gamma(\alpha + 1)},
\label{apxeq:near_zero_L}
\end{equation}
where large $n$ is also assumed, hence comparing Eqs.~\eqref{apxeq:near_zero_y} and~\eqref{apxeq:near_zero_L} we have
\begin{equation}
    \frac{C}{\Gamma(\alpha + 1)} \, n^{\alpha/2}
    = \frac{n^\alpha}{\Gamma(\alpha + 1)} 
    \Rightarrow
    C = n^{\alpha/2}.
\label{apxeq:C_coeff}
\end{equation}
hence, the full solution given by Eq.~\eqref{apxeq:start_solution} approximated for large $n$ and much smaller $x$ by Eq.~\eqref{apxeq:near_zero_y} with $C$ coefficient from Eq.~\eqref{apxeq:C_coeff} reads
\begin{equation}
    y(x) 
    = x^{-\alpha/2} \ee^{x/2} n^{\alpha/2} J_\alpha(2\sqrt{nx}),
\end{equation}
which is precisely the formula we wanted to derive.

\section{Numerical calculations for large factorials}
\label{apx:logarithmic}
For large values of $[\bar{n}]$ factorials in Eq.~\eqref{eq:ln_overlap}, employed to evaluate the semiclassical limit of spontaneous emission, are hard to calculate as written. Below, we incorporate numerical calculations based on the logarithmic expressions.

In the formula for the $\Gamma$ rate the modulus $|{}_\pm\braket{\ell}{n}_\mp|$ is used and hence we will work with it as well
\begin{equation}
    |{}_\pm\braket{\ell}{n}_\mp|
    = \sqrt{\ell! \, n!} \; \ee^{-\tfrac{1}{2}\left| \tfrac{\Omega_a}{2\omega_L} \right|^2} \sum_{k=0}^{\mathrm{min}(\ell, n)} \tfrac{(-1)^k \left| \tfrac{\Omega_a}{2\omega_L} \right|^{\ell + n -2k}}{k! (\ell - k)! (n - k)!}.
\end{equation}
The trick is to use logarithms as a more efficient way of calculating large numbers that at the end provide small results
\begin{equation}
    \begin{split}
    \ln(|{}_\pm\braket{\ell}{n}_\mp|) 
    &= -\tfrac{1}{2} \left| \tfrac{\Omega_a}{2\omega_L} \right|^2 + \tfrac{1}{2}\ln\Gamma(\ell+1) + \tfrac{1}{2} \ln\Gamma(n+1)
    \\
    &+ \ln\sum_{k=0}^{\min(\ell, n)}(-1)^k \underbrace{\tfrac{\left| \tfrac{\Omega_a}{2\omega_L} \right|^{\ell+n-2k}}{\Gamma(k+1) \, \Gamma(\ell-k+1) \, \Gamma(n-k+1)}}_{S_k}.
    \end{split}
\label{apxeq:ln_matrix_element}
\end{equation}
The sequence from the sum in Eq.~\eqref{apxeq:ln_matrix_element} can be also written as 
\begin{equation}
    \begin{split}
    \ln(S_k) 
    = (\ell + n - 2k) \ln \left| \tfrac{\Omega_a}{2\omega_L} \right| - \ln\Gamma(k+1) 
    \\
    - \ln\Gamma(\ell-k+1) - \ln(n-k+1),
    \end{split}
\label{apxeq:ln_Sk}
\end{equation}
from which we can efficiently find maximal value $\ln(S_\text{max}) = \max(\ln(S_k))$, and write the logarithm of the sum in Eq.~\eqref{apxeq:ln_matrix_element} as
\begin{equation}
    \begin{split}
    \ln\sum_{k=0}^{\min(\ell,n)} (-1)^k S_k 
    &= 
    \ln(S_\text{max}) 
    \\
    &+ \ln\sum_{k=0}^{\min(\ell,n)} (-1)^k \ee^{\ln(S_k) - \ln(S_\text{max})}.
    \end{split}
\label{apxeq:ln_sum_Sk}
\end{equation}
This form provides simpler numerical calculations, hence the full formula \eqref{apxeq:ln_matrix_element} after inserting Eqs.~\eqref{apxeq:ln_Sk} and~\eqref{apxeq:ln_sum_Sk} reads
\begin{equation}
    \begin{split}
    \ln(|{}_\pm\braket{\ell}{n}_\mp|) 
    = 
    -\tfrac{1}{2} \left| \tfrac{\Omega_a}{2\omega_L} \right|^2 + \tfrac{1}{2}\ln\Gamma(\ell+1) + \tfrac{1}{2} \ln\Gamma(n+1) 
    \\
    + \ln(S_\text{max}) + \ln\sum_{k=0}^{\min(\ell,n)} (-1)^k \ee^{\ln(S_k) - \ln(S_\text{max})}.
    \end{split}
\label{apxeq:statesOverlapingLog}
\end{equation}
In the numerical implementation the sum in Eq.~\eqref{apxeq:statesOverlapingLog} can be split into positive and negative sums to avoid direct summation of alternating sign small terms. Finally, after calculations we can go back simply using exponent
\begin{equation}
    |{}_\pm\braket{\ell}{n}_\mp| = \ee^{\ln(|{}_\pm\braket{\ell}{n}_\mp|)}.
\end{equation}

\begin{figure}
    \centering
    \includegraphics[width=0.85\linewidth]{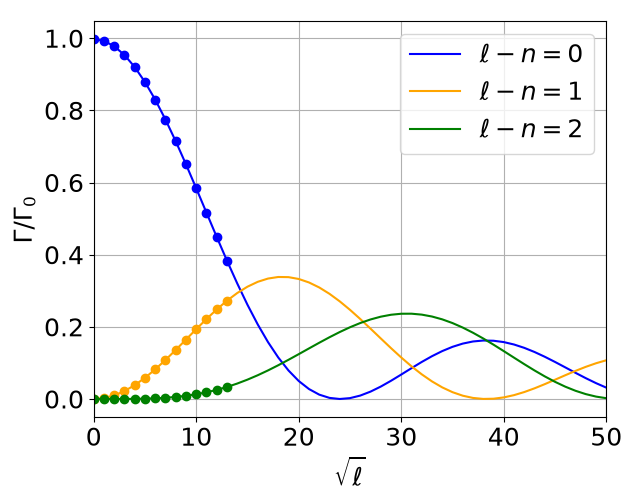}
    \caption{Absolute values of $\Gamma$ rates for different values of $n$ with use of the regular formula Eq.\eqref{eq:ln_overlap} (dots), and logarithmic formula Eq.~\eqref{apxeq:statesOverlapingLog} (solid lines) for different p = $\ell - n$ values. Parameters used: $\Omega_a/\omega_0 = 0.1$, $\omega_L/\omega_0 = 1.0$.}
    \label{fig:regVSlog}
\end{figure}

In Fig.~\ref{fig:regVSlog} we show that the regular formula for matrix element is perfectly recreated by the logarithmic one and significantly extends the range of values for $n$. For the regular formula the computational threshold for our case was around $n=170$ (dots).

%\bibliography{asym2}% Produces the bibliography via BibTeX.


\end{document}